\documentclass{kapedbk} % Computer Modern font calls

\usepackage{epsf}
\usepackage{graphicx}

\normallatexbib

\begin{document}

\articletitle{Orbital effects in manganites}

%\articlesubtitle{}

\author{Jeroen van den Brink}
\affil{Computational Materials Science and Mesa$^+$ Institute, Faculty of Applied Physics,\\
University of Twente, P.O. Box 217, 7500 AE Enschede, The Netherlands}
\email{jvdb@tn.utwente.nl}

\author{Giniyat Khaliullin}
\affil{Max-Planck-Institut f\"ur Festk\"orperforschung,\\
Heisenbergstr.\ 1, D-70569 Stuttgart, Germany}
\email{G.Khaliullin@fkf.mpg.de}

\author{Daniel Khomskii}
\affil{Laboratory of Applied and Solid State Physics, Materials Science Center,\\
University of Groningen, Nijenborgh 4, 9747 AG Groningen, The Netherlands}
\email{khomskii@phys.rug.nl}

\and %% <<== will make `and' appear in the Table of Contents. Use
     %%      before the last author is listed.

\begin{abstract}
We review some aspects related to orbital degrees of freedom in manganites. 
The Mn$^{3+}$ ions in these compounds have 
double orbital degeneracy and are strong Jahn-Teller ions, causing 
structural distortions and orbital ordering. We discuss ordering 
mechanisms and the consequences of orbital order.
The additional degeneracy of low-energy states and the extreme 
sensitivity of the chemical bonds to the spatial orientation of 
the orbitals result in a variety of competing interactions.
This quite often leads to frustration of classical ordered states 
and to the enhancement of quantum effects. Quantum fluctuations and 
related theoretical models are briefly discussed, including the 
occurence of resonating orbital bonds in the metallic phase of 
the colossal magnetoresistance manganites.  

\end{abstract}


\tableofcontents

\section{Introduction}
\label{Sec:intro}

When considering the properties of real systems with strongly
correlated electrons, such as transition metal (TM) oxides, one often
has to take into account, besides the charge and spin degrees of
freedom, described e.g. by the nondegenerate Hubbard model, also the
orbital structure of corresponding TM ions.  These orbital degrees of
freedom are especially important in cases of the so-called orbital
degeneracy --the situation when the orbital state of the TM ions in
a regular, undistorted coordination (e.g.\ in a regular
$O_6$-octahedron) turns out to be 
degenerate~\cite{goodenough,kugel,nagaosa,oles00}. This is e.g. the
situation with the ions Cu$^{2+}$ ($d^9$), Mn$^{3+}$ ($d^4$),
Cr$^{2+}$ ($d^4$), low-spin Ni$^{3+}$ ($d^7=t_{2g}^6e_g^1$). In an isolated centre 
this degeneracy gives rise to the famous Jahn-Teller
effect~\cite{jahn}, and in concentrated systems to a
cooperative transition which may be viewed as a simultaneous structural
phase transition that lifts the orbital degeneracy (cooperative
Jahn-Teller transition) and orbital ordering (OO) transition (or 
quadrupolar ordering transition, a terminology that is often used in rare
earth compounds).

\begin{figure}
\epsfxsize=60mm
\centerline{\epsffile{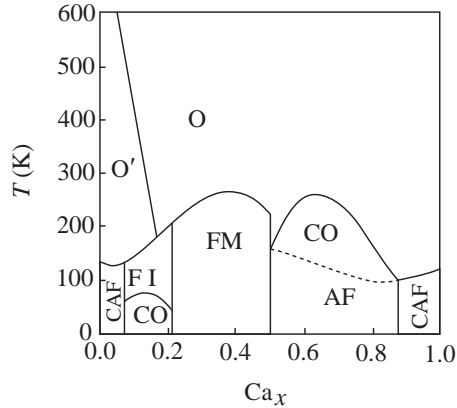}}
\caption{Phase diagram of La$_{1-x}$Ca$_x$MnO$_3$ (after S.-W. Cheong).  
O: orthorhombic phase with rotated regular octahedra, 
O$'$: orthorhombic phase with Jahn-Teller distortions, CAF: canted antiferromagnet or
phase separated state,
FI: ferromagnetic insulator, CO: charge order, FM: ferromagnetic metal, 
AF: antiferromagnet.
}
\label{Fig:phase}
\end{figure}

\begin{figure}
\noindent
\centering
\setlength{\unitlength}{0.7\linewidth}
\begin{picture}(1,0.55)
\epsfxsize=20mm
\put(-0.1,0){\epsfxsize=0.4\unitlength\epsffile{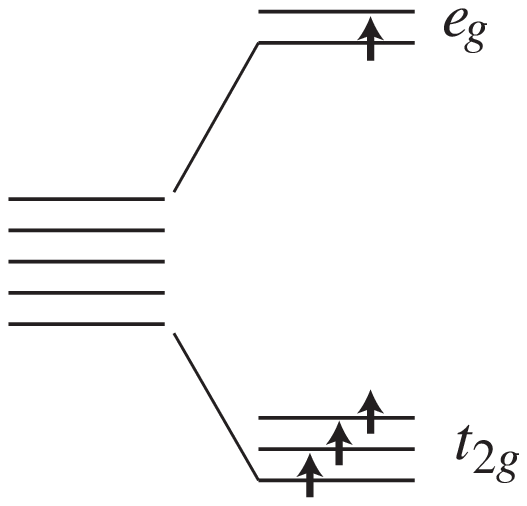}}
\epsfxsize=80mm
\put(0.4,0){\epsfxsize=0.6\unitlength\epsffile{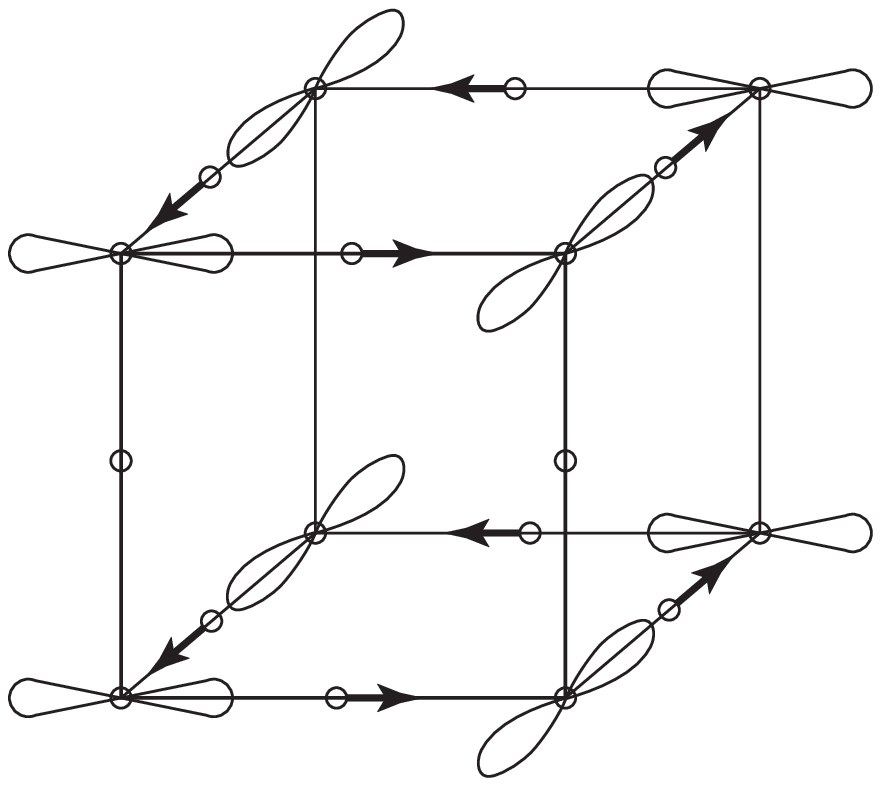}}
\end{picture}\\[10pt]
\caption{Left: splitting of $3d$-levels in a cubic crystal field
(regular MnO$_6$-octahedron).  Electron occupation of the four $3d$-electrons
in Mn$^{3+}$ is shown by arrows.
Right:  orbital ordering of LaMnO$_3$.  Arrows indicate the 
displacements of oxygen ions.}
\label{Fig:3dlevels}
\end{figure}

All these effects play a very important role in the materials which
became very popular recently --in manganites with colossal
magnetoresistance (CMR).  A typical example is the system
La$_{1-x}$Ca$_x$MnO$_3$ (there may be other rare earths, e.g. Pr, Nd,
or Bi instead of La, or other divalent cations --Sr, Ba, Pb-- instead
of Ca; there exist also layered materials of this kind).  The characteristic
phase diagram of these systems is shown schematically in
Fig.~\ref{Fig:phase}.  The  undoped material LaMnO$_3$, which is an
antiferromagnetic insulator, contains  typical Jahn-Teller ions
Mn$^{3+}$ (electronic configuration $t_{2g}^3e_g^1$) --i.e. it is
orbitally doubly-degenerate, see Fig.~\ref{Fig:3dlevels}.   Thus we can expect
that the orbital degrees of freedom may significantly influence the
properties of CMR manganites --an idea which is largely supported
by experiments.  In this  paper we review some
of the aspects of the physics of manganites connected with orbital
degrees of freedom. This field is actually
already quite large and well developed,  and  of  course  we will not be
able to cover all of it;  much
of  the  material  presented  will  be
based on the investigations in which we
ourselves  participated.   Some  of   the
general concepts  used below  are also
presented in~\cite{khomskii,khomskii2}.
There are a lot of different questions, and a wealth of properities 
(see the phase diagram in Fig.~\ref{Fig:phase}) in which orbitals are important.
In this chapter we give a general overview of these properties. Most
topics we will discuss on a rather qualitative level, trying to explain
the main physical effects, but without going into too many details.
However,  there are some very interesting and important problems
in the orbital physics of manganites, which require a more detailed
explanation. This, in particular, is the case with genuine
quantum effects which may be very important, especially (but not only)
in the "optimal doped" ferromagnetic metallic phase.
These questions are treated in Section~\ref{Sec:quantum}
in more detail and on a somewhat more theoretical level.

\subsection{Main features of phase diagram: orbital effects}
\label{Sec:phase diagram}

As already mentioned above, the undoped LaMnO$_3$ with the perovskite
structure contains strong Jahn-Teller ions Mn$^{3+}$.  They are known
to induce a rather strong local distortion in all the insulating
compounds that contain them~\cite{kugel}.  Also in LaMnO$_3$ it is well
known that there exists an orbital ordering and a concomitant lattice
distortion: $e_g$-orbitals of Mn$^{3+}$ ions are
ordered in such a way that at the neighbouring Mn sites the
alternating $3x^2-r^2$ and $3y^2-r^2$-orbitals are occupied, i.e. the local
O$_6$-octahedra are alternatingly elongated along $x$ and
$y$-directions, see Fig.~\ref{Fig:3dlevels}. Let us mention here the
typical energy scales that are involved in the manganites. 
The ferromagnetic Hunds rule exchange among $t_{2g}$ and $e_g$-electrons 
is $J_H \approx 0.8$ eV per spin pair, the crystal field splitting between these 
levels (see Fig.~\ref{Fig:3dlevels}) is $10Dq \approx 2-3$ eV, and the 
Jahn Teller energy, $E_{JT}$, which is the splitting of the $e_g$-states
by the lattice distortion, is typically 
an order of magnitude smaller than $10Dq$.
\begin{figure}
\epsfxsize=40mm
\centerline{\epsffile{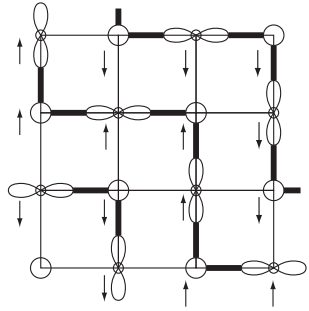}}
\caption{Charge, orbital and spin ordering in the basal
($xy$)-plane of manganates at $x=0.5$.  Arrows denote the spin ordering. 
The spin zigzags, typical for the CE-type magnetic order, 
are shown by thick lines.
}
\label{Fig:ce_phase}
\end{figure}

Orbital ordering is also known to exist in most manganites in another
well-defined region of the phase diagram --at half-doping $x=0.5$.  In
this situation with decreasing temperature charge
ordering --the checkerboard arrangement of Mn$^{3+}$ and Mn$^{4+}$
ions in the basal plane, see
Fig.~\ref{Fig:ce_phase}
~\cite{goodenough,wollan,jirak} -- sets in.   The Mn$^{3+}$-ions
with localized electrons again have an orbital degeneracy (Mn$^{4+}$ ($t_{2g}^3$)
ions are nondegenerate, Fig.~\ref{Fig:3dlevels}) and develop the orbital
ordering shown in Fig.~\ref{Fig:ce_phase}.  Both the charge ordering (CO) and
the OO occur simultaneously at the same temperature, although from
some data it follows that probably the CO is the driving force, and
the OO follows it~\cite{zimmermann}; this however is still an open
question.

According to the well-known Goodenough--Kanamori--Anderson rules (see
e.g.~\cite{goodenough,khomskii,khomskii2}) the magnitude and even the
sign of the magnetic exchange depend on the type of orbitals
that are occupied.  Thus if the  orbitals occupied
by one electron (half-filled orbitals) are directed towards each another, one has a 
strong antiferromagnetic
coupling; if however these orbitals are directed away from each other
(are mutually orthogonal) we would
have a ferromagnetic interaction.  That is why the undoped LaMnO$_3$
(Fig.~\ref{Fig:3dlevels}) has the A-type magnetic ordering --the spins in the
$(x,y)$-plane order ferromagnetically, the next $xy$-layer being
antiparallel to the first one.

There are two more regions of the phase diagram of Fig.~\ref{Fig:phase} in
which  orbital effects apparently play an important role, although
the detailed picture is less clear.  These are the low-doped region
$0.1 \leq x \leq 0.2 \ - \ 0.3$ (depending on the specific system
considered) in which one often observes the ferromagnetic insulating
(FI) and presumably charge-ordered phase.  This is the case of
La$_{1-x}$Ca$_x$MnO$_3$ ($0.1 \leq x \leq 0.25$),
La$_{1-x}$Sr$_x$MnO$_3$ close to $x=\frac18$ ($0.1\leq x\leq0.18$)~\cite{klingerer,endoh} and
Pr$_{1-x}$Ca$_x$MnO$_3$ ($0.15\leq x\leq0.3$)~\cite{jirak}.

It is rather uncommon to have a FI state: typically insulating
materials of this class are antiferromagnetic, and ferromagnetism goes
hand in hand with metallicity, which is naturally explained in
the model of double exchange~\cite{degennes}.  The only possibility
to obtain the FI state in perovkites is due to a certain particular
orbital ordering favourable for ferromagnetism~\cite{khomskii} (the FI
state can appear also in systems in which there exists the
$90^\circ$-superexchange: the TM--O--TM angle is close to
$90^\circ$).  But what is the detailed ordering in this low-doped
region, is not completely clear, see Section~\ref{Sec:lowdoping}.

Another interesting, and much less explored, region is the overdoped
manganites, $x>0.5$.  Typically in this case we have an insulating
state, sometimes with the CO and OO state in the form of
stripes~\cite{radaelli} or bistripes~\cite{mori}.
The choice between these two options is still a  matter of
controversy (see e.g.~\cite{radaelli,wang}), as well as the detailed type of magnetic
ordering in this case.  We will return to this point in section~\ref{Sec:x.gt.half}.
When discussing the properties of overdoped manganites at $x>0.5$,
one should mention an important fact: the very strong asymmetry of the
typical phase diagram of manganites.  As seen e.g.\ from
Fig.~\ref{Fig:phase}, there usually exists a rather large ferromagnetic
metallic region (FM) for $x<0.5$, but almost never for $x>0.5$ (only
rarely one observes bad metallic behavior and unsaturated ferromagnetism in a
narrow concentration range in some overdoped
manganites~\cite{martin}).  However from the standard
double-exchange model one can expect the appearance of a FM phase not
only in hole-doped LaMnO$_3$ ($x<0.5$) but in electron-doped
CaMnO$_3$ ($x>0.5$) as well. Orbital degeneracy may play some
role in explaining this asymmetry~\cite{vandenbrink2} --see section~\ref{Sec:x.ge.half}.

There exists also a problem as to what are the orbitals doing in the optimally
doped ferromagnetic and metallic manganites. Usually one completely
ignores orbital degrees of freedom in this regime, at least at low
temperatures; this is supported by the experimental observations
that the MnO$_6$-octahedra are completely regular in this case.
But orbital degrees of freedom cannot just vanish. In this
doping regime there is a strong competition between the tendency of 
orbitals to order locally and the kinetic energy of the charge carriers 
that tends to destroy long range orbital order. This is comparable to 
the situation in High T$_c$ superconductors, where the long-range 
antiferromagnetic order of, in this case, spins is frustrated by
mobile charge carriers. In an analogous way the mobile holes in optimally 
doped manganites can melt the orbital order: 
in the ferromagnetic metallic phase of the manganites the orbital degrees of 
freedom do not simply disappear, but instead the orbitals are "rotating" 
very fast and may form resonating orbital bonds, see Section~\ref{Sec:quantum}.
Another option --orbital ordering of a new type with "complex orbitals" is discussed 
in Section~\ref{Sec:complexOO}.

\subsection{Mechanisms for Orbital Ordering}
\label{Sec:OO}

Before discussing particular situations in different doping
ranges, it is worthwhile to address briefly the general question of
possible interactions of degenerate orbitals which can
lead to orbital ordering.  In transition metal compounds there are
essentially two such mechanisms.  The first one is connected with the
Jahn-Teller interaction of degenerate orbitals with the lattice
distortions, see e.g.~\cite{gehring}.  Another mechanism was proposed
in 1972~\cite{kugel2}, see also~\cite{kugel}, and is a direct 
generalization of the usual
superexchange~\cite{anderson} to
the case of orbital degeneracy.

\begin{figure}
\centering  \includegraphics[totalheight=50mm]{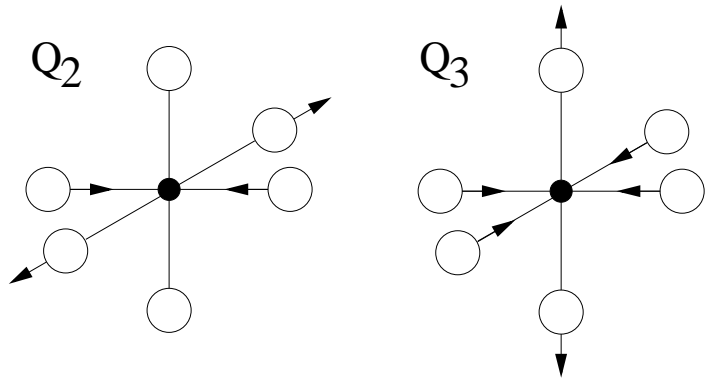}
\caption{
Schematic representation of the $Q_2$ and $Q_3$ Jahn-Teller distortions of a
MnO$_6$ octahedron. 
}
\label{Fig:Q2-Q3}
\end{figure}

A convenient mathematical way to describe orbital ordering is to
introduce operators $T_i$ of the pseudospin $\frac12$,
describing the orbital occupation, so that e.g.\ the state
$|T^z=\frac12\rangle$ corresponds to the occupied orbital
$|3z^2-r^2\rangle$, and $|T^z=-\frac12\rangle$ to $|x^2-y^2\rangle$.
The first one corresponds to a local elongation of the
O$_6$-octahedra (distortion coordinate $Q_3>0$), see Fig.~\ref{Fig:Q2-Q3}, 
and the second --to local
contraction $Q_3<0$~\cite{kanamori}.  The second degenerate
$E_g$-phonon which can also lift electronic $e_g$-degeneracy, $Q_2$, 
see Fig.~\ref{Fig:Q2-Q3},
corresponds to a pseudospin operator $T^x$.  One can describe an
arbitrary distortion and corresponding wave function by  linear
superpositions of the states $|T^z=+\frac12\rangle$ and
$|T^z=-\frac12\rangle$:
\begin{equation}
\textstyle|\theta\rangle=\cos\frac\theta2|\frac12\rangle+\sin\frac\theta2|-\frac12\rangle
\label{eq1}
\end{equation}
where $\theta$ is an angle in $(T^z,T^x)$-plane.

The  Jahn-Teller mechanism for orbital ordering starts from
the electron--phonon interaction, which in our case can be written in
the form
\begin{equation}
H=\sum_{i q} g_{iq}[T^z_i(b^\dagger_{3q}+b^{\vphantom{\dagger}}_{3,-q})+
T_i^x(b^\dagger_{2q}+b^{\vphantom{\dagger}}_{2,-q})]+
\sum_{\alpha q} \omega_{\alpha q}b^\dagger_{\alpha q}b^{\vphantom{\dagger}}_{\alpha q}
\label{Hpho_orb}
\end{equation}
where $\alpha=2;3$ and $b^\dagger_3$ and $b^\dagger_2$ are the phonon
operators corresponding to $Q_3$ and $Q_2$ local modes.  Excluding
the phonons by a standard procedure, one obtains the orbital
interaction having the form of a pseudospin--pseudospin interaction
\begin{equation}
H_{\it eff}=\sum_{ij}J_{ij}^{\mu\nu}T_i^\mu T_j^\nu
\label{eq3}
\end{equation}
where
\begin{equation}
J_{ij}\sim\sum_q\frac{g_q^2}{\omega_q}e^{iq(R_i-R_j)}
\end{equation}
and $\mu,\nu=x,z$.  Due to different dispersion of the different
relevant phonon modes, and due to the anisotropic nature of
electron--phonon coupling, the interaction (\ref{eq3}) is in general
anisotropic.

Similarly, the exchange mechanism of orbital ordering may be
described by the Hamiltonian containing the pseudospins $T_i$, but it contains
also the ordinary spins $\vec S_i$.  The effective superexchange Hamiltonian 
can be derived starting from the
degenerate Hubbard model~\cite{kugel2}, and it has schematically the
form
\begin{equation}
H=\sum_{ij}\bigl\{J_1\vec S_i\vec S_j+J_2(T_iT_j)+J_3(\vec S_i\vec S_j)(T_iT_j)\bigr\}.
\label{eq5}
\end{equation}
Here the orbital part $(T_iT_j)$, similar to (\ref{eq3}), is in
general anisotropic, whereas the spin exchange is Heisenberg-like.
In contrast to the Jahn-Teller induced interaction, the exchange
mechanism describes not only the orbital and spin orderings
separately, but also the coupling between them (last term
in~(\ref{eq5})).  This mechanism is rather successful in explaining
the spin and orbital structure in a number of
materials~\cite{kugel,kugel2}, including LaMnO$_3$ (for the latter system
one has to invoke also the anharmonicity effects~\cite{kugel2} --see
also~\cite{khomskii2p}).

As to the electron-lattice interaction, typically one includes
mostly the coupling with the
local --i.e. optical-- vibrations~\cite{halperin}.  However no less
important may be the interaction with the long-wavelength acoustical
phonons, or, simply speaking, with the elastic deformations.
Generally, when one puts an impurity in a crystal, e.g.\
replacing the small Mn$^{4+}$ ion in CaMnO$_3$ by the somewhat larger
Mn$^{3+}$ ion, which in addition causes a local lattice distortion due
to the Jahn-Teller effect (i.e.\ we replace a ``spherical'' Mn$^{4+}$ ion
by an ``ellipsoidal'' Mn$^{3+}$), this creates a strain field which is
in general anisotropic and decays rather slowly, as
$1/R^3$~\cite{eshelby,khachaturyan}.  A second ``impurity'' of this
kind ``feels'' this strain, which leads to an effective long-range
interaction between them.  This can naturally lead to the spontaneous
formation of different superstructures in doped
materials~\cite{khomskii3,khomskii4}.  Thus, there may appear vertical or
diagonal stripes, even for non-Jahn-Teller systems.  In case of
manganites one can show that there appears e.g. an effective attraction
between $3x^2-r^2$ and $3y^2-r^2$-orbitals in $x$ and $y$-direction; this
immediately gives the orbital ordering of LaMnO$_3$-type shown in
Fig.~\ref{Fig:3dlevels}.  For $x=0.5$, assuming the checkerboard charge ordering, one
gets from this mechanism the correct orbital ordering shown in
Fig.~\ref{Fig:ce_phase}~\cite{khomskii3,khomskii4}.  And for overdoped manganites one can
get either single or paired stripes, depending on the ratio of
corresponding constants:  One can show~\cite{halperin,khomskii3,khomskii4} that
for a diagonal pair like the ones in Fig.~\ref{Fig:stripes}, one gets an attraction
of the same orbitals $3x^2-r^2$ and $3x^2-r^2$ or $3y^2-r^2$ and $3y^2-r^2$, but
repulsion of $3x^2-r^2$ and $3y^2-r^2$. Thus, if one
takes into account only these nearest neighbour diagonal
interactions, the single stripe phase of Fig.~\ref{Fig:stripes} would be more
favourable than the paired stripes of the Fig.~\ref{Fig:stripes}. However the
latter may in principle be stabilizes by more distant interactions
like those for a pair of Mn$^{3+}$ ions along $x$ and $y$-directions
in Fig.~\ref{Fig:stripes}. Which state is finally more favourable, is still not
clear, see Ref.~\cite{khomskii4}.

\section{Orbital order for $x < 1/2$}
\label{Sec:x.lt.half}

In the previous section we have given some examples that show that the
interaction of orbital degrees of freedom among themselves and the interaction
with electron spins or with the lattice, can give rise to long-range 
orbital ordered states. Experimentally the manganites seem to be especially
susceptible to an orbital order instability at commensurate doping 
concentrations: the undoped system is a canonical example of an 
orbitally-ordered Mott insulator and for $x=1/2$ orbitals are ordered in most manganites.
Below we discuss the situation for $x=1/8$ and $x=1/4$ and the possibility
of orbital order in the metallic phase.

Every ordered state has one or more elementary excitations related to 
the actual symmetry that is broken by the long-range order.
In spin systems this symmetry is often continuous, which leads to the
occurence of a Goldstone mode, in this case a spin-wave --magnon-- that has
a vanishing excitation energy when its wavelength is very long.
Related to the ordering of orbitals there should therefore be an elementary
excitation with orbital character: the orbiton. Recently this orbital wave 
was actually observed experimentally~\cite{saitoh}. In this and the next section
we discuss some of its properties, emphasizing that the orbital mode is
gapped and that it strongly interacts with lattice and spin degrees 
of freedom~\cite{brink01}.

\subsection{Undoped and lightly doped manganites}
\label{Sec:lowdoping}

As already mentioned in section~\ref{Sec:intro}, typically there exist a 
ferromagnetic
insulating region at low doping
($0.1 \stackrel{<}{\scriptstyle \sim} x   \stackrel{<}{\scriptstyle \sim} 0.18$
for the LaSr system, $x \stackrel{<}{\scriptstyle \sim} 0.25$ for LaCa, $0.15 
 \stackrel{<}{\scriptstyle \sim}  x  \stackrel{<}{\scriptstyle \sim} 0.3$ for PrCa). The
problem is to explain the origin of the FI state in this case.
Apparently it should be connected with an orbital ordering of some
kind; but what is the specific type of this ordering, is largely
unknown.

The most complete, but still controversial, data exist on 
the LaSr-system close to $x=1/8$. There
exists a superstructure in this system~\cite{yamada,klingerer}, and an orbital
ordering was detected in the FI phase in~\cite{endoh}. Certain
orbital superstructures were also seen by the resonant X-ray
scattering in  Pr$_{0.75}$Ca$_{0.25}$MnO$_3$~\cite{zimmermann2}. Both
these systems, however, were looked at by this method only at one $k$-point [300],
which is not sufficient to uniquely determine the type of orbital
ordering. The existing structural data, or rather the interpretation of this data, 
is still controversial~\cite{Inami99,Yamada00,Cox01}.

\begin{figure}
\epsfxsize=65mm
\centerline{\epsffile{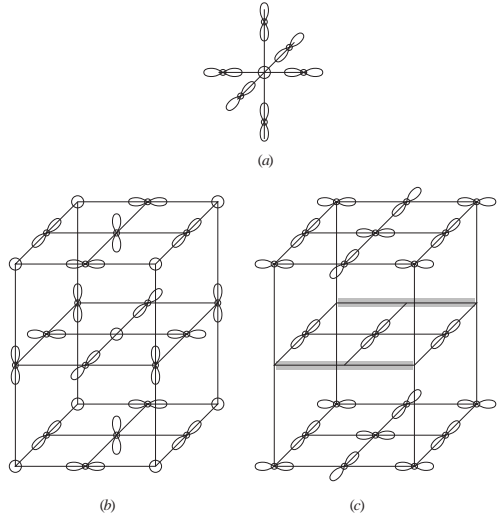}}
\caption{Orbital polarons and possible types of orbital
ordering in low doped manganites:
($a$) Orbital polaron close to a Mn$^{4+}$ ion; ($b$)~Ordering of
orbital polarons for $x=0.25$ in a bcc-lattice; ($c$)~An alternative
charge and orbital ordering, obtained for $x=1/8$ in~\cite{mizokawa}.  Notations
are the same as in Figs.~\ref{Fig:ce_phase} and \ref{Fig:stripes}.  
Shaded lines--``stripes'' containing holes.}
\label{Fig:polaron}
\end{figure}

Theoretically two possibilities were discussed in the
literature~\cite{mizokawa,mizokawa2}. First of all one can
argue that when one puts a Mn$^{4+}$ ion into Mn$^{3+}$ matrix, the
orbitals of all the ions surrounding the localized hole (Mn$^{4+}$)
would be directed towards it, see Fig.~\ref{Fig:polaron}$a$~\cite{mizokawa2,khaliullin}.
One can call such  state an orbital polaron.
We use the concept of 'orbital polaron' in a rather loose sense here and for a
somewhat more formal discussion we refer to Section~\ref{Sec:polarons}.
These polarons, which according to
Goodenough--Kanamori--Anderson rules would be ferromagnetic, can then
order e.g.\ as shown in Fig.~\ref{Fig:polaron}$b$ for $x=0.25$. The calculations
carried out in \cite{mizokawa2} show that this state is indeed
stable, and it corresponds to a ferromagnetic insulator. Thus it is a
possible candidate for a FI state at $x\simeq1/4$ e.g.\ in Pr--Ca
system.

However there exist an alternative possibility. Calculations
show~\cite{mizokawa} that a similar state with ordered polarons is also
locally stable for  $x\simeq1/8$. But it turned out that the lower
energy is reached in this case by different type of charge and
orbital ordering, Fig.~\ref{Fig:polaron}$c$~\cite{mizokawa}: the holes are localized
only in every second $xy$-plane, so that one such plane containing
only Mn$^{3+}$ ions develops the orbital ordering of the type of
LaMnO$_3$, Fig.~\ref{Fig:3dlevels}, and the holes in the next plane concentrate in
``stripes'', e.g.\ along $x$-direction. This state also turns out to be
ferromagnetic, and the superstructure obtained agrees with the
experimental results of \cite{yamada} and \cite{klingerer} for
La$_{1-x}$Sr$_x$MnO$_3$, $x\simeq1/8$. One can think that the
situation can be also similar for $x\simeq1/4$ which would agree
with the data of~\cite{zimmermann2}. This type of the charge ordering
(segregation of holes in every second plane) may be favourable due
to an extra stability of the LaMnO$_3$-type orbital ordering,
strongly favoured by the elastic interactions~\cite{khomskii4}, 
as discussed in section~\ref{Sec:intro}. A "mixed" possibility is in 
principle also possible: there may exist for example a similar 
charge segregation into a hole-rich and hole-poor plane, but the orbitals
may rather behave as is shown in Fig. 1.5$a$, i.e. contrary to fig 1.5$c$,
there may be partial occupation of $3z^2-r^2$-orbitals at certain sites.

\subsection{Possible orbital ordering in metallic phase}
\label{Sec:complexOO}
Let us discuss the most important phase --that of optimal
doping, $x\simeq0.3$ - $0.5$. In most cases the systems in this
doping range  are ferromagnetic and metallic at low temperatures,
although the residual resistivity is usually relatively large.

Now, the question is: what are the orbitals doing in this phase?
Experimentally  one observes that the macroscopic Jahn-Teller ordering
and corresponding lattice distortion is gone in this regime. La--Ca
system remains orthorhombic in this concentration range, but that is due
to the tilting of the O$_6$ octahedra, the octahedra themselves being
regular (all the Mn--O distances are the same). The structure of the
La--Sr manganites in this regime is rhombohedral, but again all the
Mn--O distances are equal. Moreover even the local probes such as
EXAFS or PDF (pair distribution function analysis of
neutron scattering)~\cite{louca}, which detect local
distortions above and close to $T_c$, show that for $T\to0$ they
completely disappear, and MnO$_6$-octahedra are regular even locally.

What happens then with the orbital degrees of freedom? There are
several possibilities. One is that in this phase the system may
already be an ordinary metal, the electronic structure of which is
reasonably well described by the conventional band theory. In this
case we should not worry about orbitals at all: we may have a band
structure consisting of several bands, some of which, not necessarily
one, may cross  Fermi-level, and we should not speak of orbital
ordering in this case, just as we do not use this terminology
and do not worry about orbital ordering in metals like Al or Nb which
often have several bands at the Fermi-level.

If however there exist strong electron correlations in our system
(i.e. the Hubbard's on-site repulsion $U$ is bigger that the
corresponding bandwidth) --one should worry about it. The orbital
degrees of freedom should then do
something. In principle there exists two options. One is that the ground
state would still be disordered due to quantum fluctuations, forming
an orbital liquid~\cite{ishihara97}, similar in spirit to the RVB state of the spin
system (we can speak of the pseudospin RVB state). We discuss this physics
in the next section.

There is also an alternative possibility: there
may in principle occur an orbital ordering of a novel
type, without any lattice distortion, involving not the orbitals of
the type~(\ref{eq1}), but the {\it complex} orbitals --linear
superpositions of the basic orbitals $3z^2-r^2$ and $(x^2 - y^2)$ with the
complex coefficients, e.g.
\begin{equation}
|\pm\rangle=\frac{1}{\sqrt2}\Bigl(|3z^2-r^2\rangle\pm
i|x^2-y^2\rangle\Bigr).
%\label{eq7}
\end{equation}
This possibility was first suggested in~\cite{khomskii5}
and then explored in~\cite{vandenbrink3,maezono}; independently a similar conclusion
was reached a bit later in~\cite{takahashi}.
One may easily see that the 
distribution of the electron density in this state is the same in all three
directions, $x$, $y$ and~$z$. Thus this ordering does not induce any
lattice distortion --the MnO$_6$ octahedra remain regular, and the
system is cubic (if we ignore tilting of the octahedra). On the other
hand, the state with complex coefficients, as always is the case with complex wave
functions, breaks time-reversal invariance, i.e. this state is in
some sense magnetic. One can show however that the magnetic dipole
moment in this case is zero --it is well known that the orbital
moment is quenched in $e_g$-states (these states are actually
$|l^z=0\rangle$ and $\frac{1}{\sqrt2}(|2\rangle + \mathopen{|}-2\rangle)$
states of the $l=2$ $d$-orbitals). Similarly,
the magnetic quadrupole moment is also zero, by parity arguments. The
first nonzero moment in this state is a magnetic octupole. Indeed,
the actual order parameter in this case is the average
%\begin{equation}
$
\eta=\langle M_{xyz}\rangle=\langle{\rm S}L_xL_yL_z\rangle\neq0
$
%\label{eq8}
%\end{equation}
where $l_\alpha$ are the components of the orbital moment $l=2$ of
$d$-electrons, and S means the symmetrization. This operator is
actually proportional to the $T^y$-operator of pseudospin, i.e. the
order parameter of this type of orbital ordering is 
$\eta=\langle T^y\rangle$.
One can visualize this state as the one in which there exist orbital
currents at each unit cell. 

\subsection{Orbitons: orbital excitations}
\label{Sec:excitations}

Each time we have certain ordering in
solids, corresponding excitations should appear. 
The quadrupolar charge ordering --orbital ordering-- should 
give rise to elementary excitations with orbital signature, as this order causes 
a breaking of symmetry in the orbital sector. 
These excitations --we may call them orbitons-- 
were first discussed shortly in~\cite{kugel,kugel2} and recently were studied
theoretically in several papers, e.g.
in~\cite{ishihara3,vandenbrink4,KHA97,brink01}. One of the problems that 
could complicate an experimental observation of these
excitations, is the usually rather strong Jahn-Teller
coupling of orbital degrees of freedom with the lattice distortions.
This could make it very difficult, if not impossible, to
``decouple'' orbitons from phonons. And indeed the experimental
efforts to observe orbitons were unsuccessful for many years. A
breakthrough came only recently when the group of Y.~Tokura
managed to observe the manifestations of orbital excitations in Raman 
scattering on
untwinned single crystals of LaMnO$_3$~\cite{saitoh}. 

The observed orbitons were interpreted by the authors as being due to electron 
correlations~\cite{saitoh}, 
but in the comment that accompagnied this publication it was immediately
mentioned that here the coupling to phonons can also be very important~\cite{Allen01}. 
Let us briefly discuss the question of the origin of orbitons; this is also 
important as they, in turn, have a large effect on spin~\cite{KHA97,FEI97} 
and charge excitations~\cite{Perebeinos00,KIL99,BHO00}, as we discussed in the previous section.

The physical aspects of the coupling between the orbital excitation and Jahn-Teller
phonons can be illustrated by considering the orbital and phonon excitations
as dispersionless. We can view this as a reasonable first approximation because orbital 
excitations 
are always gapped (see next section) and Jahn-Teller lattice excitations are optical phonons.
Applying this simplification to Eqs.(\ref{Hpho_orb}) and (\ref{eq3}) leads to the 
single-site Hamiltonian~\cite{brink01}
\begin{eqnarray}
H_{loc} &=&
[\bar{J} + 2g ( b^{\dagger}_{3} + b_{3} )] q^{\dagger} q + 
\omega_0 (b_{3}^{\dagger}b_{3}+b_{2}^{\dagger}b_{2}) \nonumber \\
&+& g (q^{\dagger}+q)  (b_{2}^{\dagger}+b_{2}), 
\label{H_loc}
\end{eqnarray}
where the $q$ operator describes an orbiton excitation, $b_{2,3}$ are the $Q_{2,3}$
phonon modes and
$
\bar{J}= 3J+4g^2/\omega_0,
$
where $g$ is the electron-phonon coupling constant, $\omega_0$ the JT phonon frequency
and $J$ the superexchange energy.
Let us discuss three important consequences of the orbiton-phonon coupling 
in  Eq.~(\ref{H_loc}). First, the coupling to the lattice
moves the orbiton to higher energy an amount $4g^2/\omega_0$. 
This shift has a straightforward physical meaning:
it is the phonon contribution to the crystal-field splitting of the $e_g$-states
caused by the static Jahn-Teller lattice deformation.
The effective orbital excitation energy is 
the sum of the local orbital exchange energy and static phonon contribution
to the crystal-field splitting.

If, however, an orbital excitation is made, it strongly interacts
with the $Q_3$ phonon, so that the orbital excitation 
can be dynamically screened by the Jahn-Teller phonons and lowered in energy.
The crystal-field splitting and screening are strongly competing as
both are governed by the energy scale set by the electron-phonon coupling. 
Finally, the orbital and $Q_2$ phonon modes mix, 
as is clear from the last term of $H_{loc}$. This implies that
the true eigenmodes of the coupled orbital-phonon system have both 
orbital and phonon character. 

\begin{figure}
\epsfxsize=90mm
\centerline{\epsffile{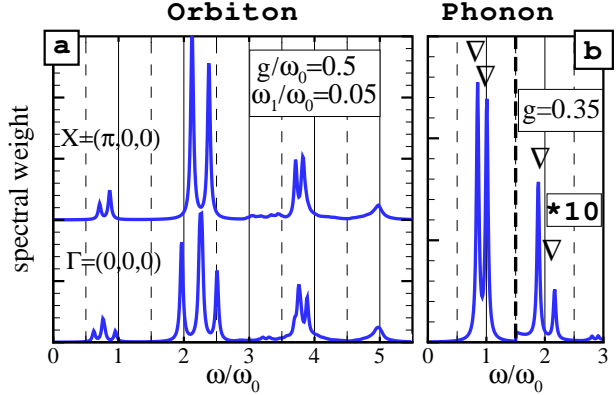}}
\vspace*{2mm}
\caption{
(a) Orbiton spectral function at the $\Gamma$ and X-point, $g=\omega_0/2$. 
(b) Spectrum of the Raman-active A$_{\rm g}$ and B$_{\rm 1g}$ phonon modes for
$g/ \omega_0=0.35$.
The experimental peak positions are indicated by $\nabla$. For $\omega > \omega_0$
the spectral weight is multiplied by 10, see Ref.~\cite{brink01}.
}
\label{Fig:spec}
\end{figure}

In general, the mixing of orbital and phonon mode gives rise to extra 
phonon satellites in the orbiton spectral function at energy intervals of $\omega_0$. 
Vice versa, due to this mixing, a low intensity orbital satellites at $\approx 3J$ 
will be present in the $Q_2$ phonon spectral function.
As discussed above, the most plausible estimates of the Jahn-Teller energy
($E_{JT}= 4 g^2/\omega_0$) are $E_{JT} \approx 200-300$ meV, and the superexchange
$J \approx 40$ meV. With these estimates, keeping in mind the
results described above --the strong mixing of the orbital excitation and
the phonons-- one comes to the conclusion that the features at 150 meV
observed in the Raman experiment~\cite{saitoh}, interpreted there as
pure orbital excitations, are rather the orbiton-derived satellites in
the phonon spectral function, see Fig.~\ref{Fig:spec}~\cite{brink01}.
A framework beyond the toy model described above is needed to establish the relevance 
of such an interpretation, and to establish the exact nature of the satellites in the
Raman spectra: this is still an open issue, both experimentally and theoretically.
But it is clear that the elementary excitations of an orbital ordered system are 
mixed modes with both orbital and phonon character, or, in other words, are
determined by both electron correlation effects and the electron-lattice interaction.

\section{Quantum effects; optimal doping}
\label{Sec:quantum}
Below the cooperative orbital/Jahn-Teller transition temperature, 
a long-range coherence of orbital polarization sets in, and 
spin-exchange interactions on every bond are fixed by the
Goodenough-Kanamori rules \cite{goodenough,khomskii,khomskii2,GOO55}. 
Implicit in this picture is that the orbital
splittings are large enough so that we can consider orbital populations
as classical numbers. Such a classical treatment of orbitals is 
certainly justified when orbital order is driven by strong cooperative 
lattice distortions that lead to a large splitting of the --initially
degenerate--
orbital levels. In this limit the orbital excitations are  
more or less localized high-energy quadrupole moment (or crystal-field) 
transitions, and
therefore they effectively just renormalize spin degrees of freedom but 
otherwise do not
effect much the physical properties of the system at low energy scales.
 
Quantum effects, however, can start to dominate the ground state properties
and elementary excitations when classical order is frustrated by 
some interaction that opposes the tendency of the orbitals to order. 
The reason that this might very well be the case in some regions 
of the phase diagram of the manganites is that orbitals strongly interact
with spins via the superexchange and that orbital order is frustrated
by the kinetic energy of carriers in a metallic system.
But the actual situation in manganites is still under debate; while in the 
undoped case
the orbital excitations are rather high in energy and therefore quite
decoupled from the spins, there are many indications that orbital dynamics 
play an essential role in the physics of doped manganites. 

In this Section we discuss  several physical examples and some theoretical models that 
show that there is a strong dynamical interplay between orbital fluctuations and spin
and charge degrees of freedom.
Both spin exchange and charge
motion are highly sensitive to the orbital bonds, and the basic idea is
that the interaction energy 
cannot be optimized simultaneously in all the bonds: this leads to the
peculiar frustrations and quantum resonances among orbital bonds. 
Of course a classical treatment in such cases gives 
very poor estimates of energies, and the 
quantum dynamics of the coupled orbital-spin-charge system 
becomes of crucial importance, as we illustrate below in the context of several 
theoretical models with coupled orbital, spin and charge degrees of freedom. 
Thus the important physics, quantum effects, largely depend on some particular
properties of orbital systems, notably the frustration in the orbital sector, 
which are absent in canonical spin and spin-charge models, e.g. the t-J model.
To explain this, we devote some time to a more theoretical discussion,
treating some orbital and spin-orbital models (Section~\ref{Sec:models} 
and~\ref{Sec:orbitalmodels}); later on in this chapter we will apply these concepts
to the discussion of properties of manganites in the cubic metallic 
phase (Sections~\ref{Sec:liquid} and~\ref{Sec:magnon})

%%%%%%%%%%%%%%%%%% Kugel-Khomskii %%%%%%%%%%%%%

\subsection{Superexchange and spin-orbital models}
\label{Sec:models}
Quantum fluctuations of the orbitals originate mainly from two 
kind of interactions. First we discuss the superexchange interaction, and
the second mechanism --the very effective frustration of orbitals by doped
holes-- we describe later on. 

Let us take as an example a toy version of the full superexchange model (proposed 
by Kugel and Khomskii~\cite{kugel}). On a three-dimensional cubic lattice 
it takes the form:
\begin{eqnarray}
H &=& \sum_{\langle i,j \rangle } \hat J^{(\gamma)}_{ij}
( \vec S_i \vec S_j + \frac{1}{4}),
\label{HAM2}\\
\hat J^{(\gamma)}_{ij} &=& J (T_{i}^{(\gamma)} T_{j}^{(\gamma)} -
\frac{1}{2}T_{i}^{(\gamma)}- \frac{1}{2}T_{j}^{(\gamma)} +
\frac{1}{4})\nonumber,
\end{eqnarray}
with $J=4 t^2/U$, where $t$ and $U$ are, respectively, the hopping integral and on-site Coulomb 
repulsion
in the Hubbard model for two-fold degenerate $e_g$-electrons at half filling. 
The structure of $T_i^{(\gamma)}$ depends 
on the index $\gamma$ which
specifies the orientation of the bond $\langle ij \rangle$ relative to
the cubic axes $a,b$ and $c$:
\begin{equation}
T_{i}^{(a/b)}=\frac{1}{4}
(-\sigma_{i}^{z}\pm \sqrt{3}\sigma_{i}^{x}),\;\;\;\;\;
T_{i}^{(c)}=\frac{1}{2}\sigma_{i}^{z},
\label{PSE}
\end{equation}
where $\sigma^z$ and $\sigma^x$ are the Pauli matrices. Physically
the $T$ operators describe the dependence of the spin-exchange 
interaction on the orbitals that are occupied,
and the main feature of this model --as is suggested by the
very form of Hamiltonian (\ref{HAM2})-- is the strong interplay between
spin and orbital degrees of freedom. It was recognized first 
in Ref.~\cite{FEI97} that this simple
model contains rather nontrivial physics: the classical
N\'eel state in Eq.~(\ref{HAM2})
(where $\langle \vec{S}_i \vec{S}_j \rangle = -1/4$) is
infinitely degenerate in the orbital sector;
this extra degeneracy must be lifted by some mechanism.
%%%%%%%%%%%%%%%%%%%%%%%%%
\begin{figure}
\centering
\setlength{\unitlength}{0.5\linewidth}
\begin{picture}(1,1)
\put(0,0){
%\epsffile{khaliullin_fig1a.eps,width=\unitlength}
\epsfxsize=\unitlength\epsffile{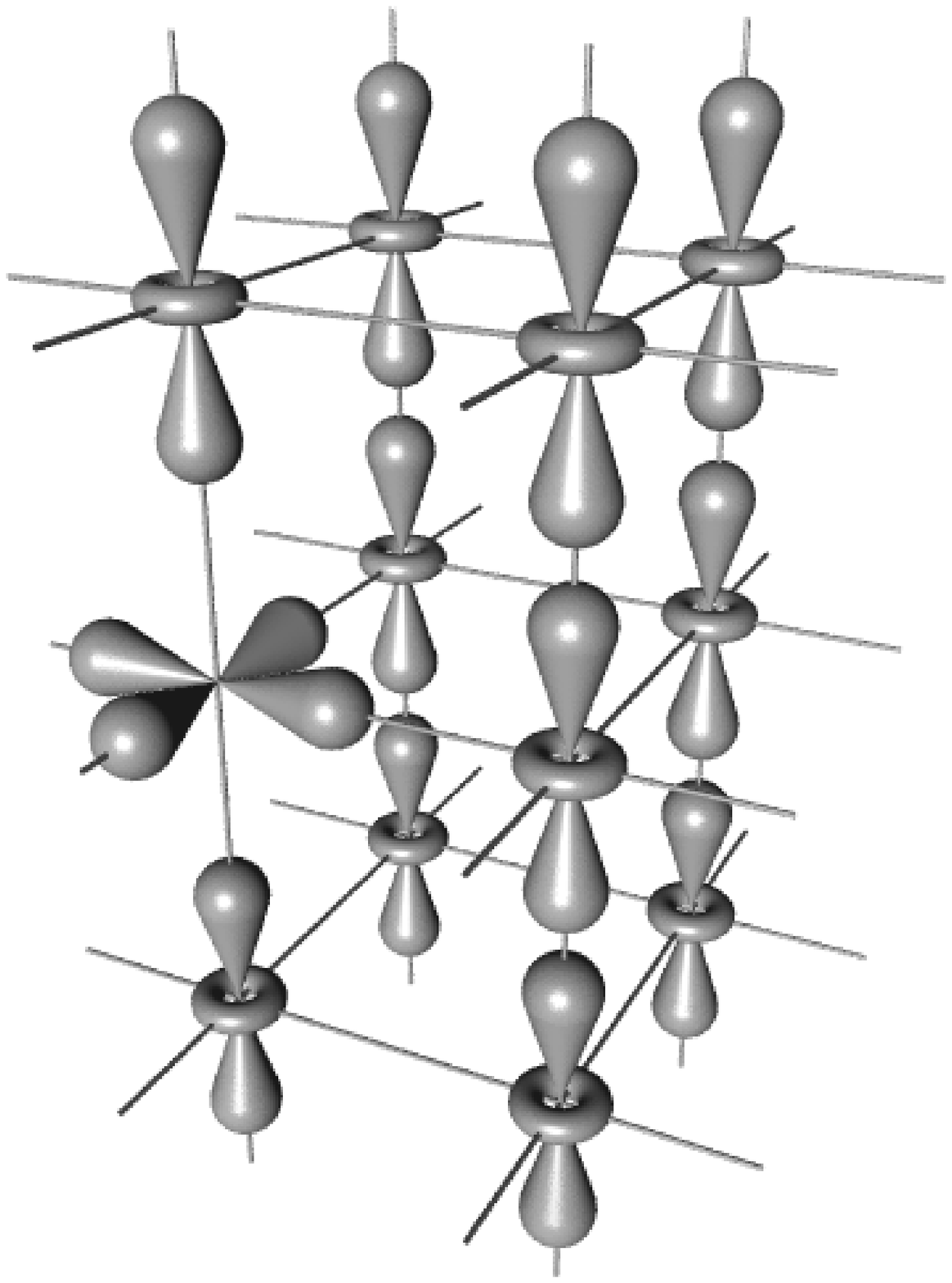}}
\put(0.1,0.53){
%\epsffile={khaliullin_fig1b.eps,width=0.11\unitlength}
\epsfxsize=0.11\unitlength\epsffile{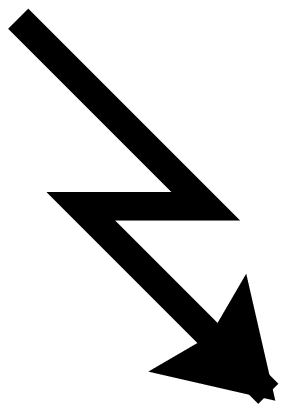}}
\end{picture}
\caption{$|3z^2-r^2\rangle$-orbital order which leads to weakly
coupled AF spin chains ($J_c=J$, $J_{\perp} = J/16$).
As discussed in Refs.\ \protect\cite{KHA97,KHA99}, this type of orbital 
ordering provides the largest energy gain due to quantum spin 
fluctuations. An orbital flip (indicated by an arrow)
modulates the strength of the neighboring exchange bonds,
breaking the $c$ chain. In the classical N\'eel state, 
such orbital excitations cost no energy~\cite{FEI97,FEI98}. 
Due to the presence of strong quasi one-dimensional spin 
fluctuations, however, a finite orbital gap opens through
the order from disorder mechanism, 
thus stabilizing this structure.}
\label{FIG:ORB}
\end{figure}
%%%%%%%%%%%%%%%%%%%%%%%%%%%% 
Before discussing this mechanism, let us first elaborate on this 
degeneracy and the importance of quantum effects in this case.
We first notice that the effective spin exchange constant in 
this model is definite positive ($\langle J^{(\gamma)}_{ij} \rangle \ge 0$)
for any configuration of orbitals, where its value can vary from 
zero to $J$, depending on the orientation of orbital pseudospins. 
We therefore expect a simple two-sublattice antiferromagnetic, G-type, 
spin order. 
There is however a problem: a classical G-type ordering has
cubic symmetry and can therefore not lift the orbital degeneracy, not even locally.
In more formal terms, the spin part $({\vec S}_i{\vec S}_j +1/4)$ of the 
Hamiltonian (\ref{HAM2}), in mean field approximation, 
simply becomes zero in this state for all bonds, 
so that these orbitals effectively do not interact --they are completely
uncorrelated-- and hence retain 
full rotational freedom on every lattice site. 
In other words, we gain no energy
from the orbital interactions that are present in the model.
This shows that from the point of view of the orbitals the classical N\'eel state is 
energetically a very poor state of the system.

The mechanism for developing intersite orbital correlations (and hence to gain energy 
from orbital ordering) must involve a strong deviation in the spin configuration from the  
N\'eel state --a deviation from $\langle \vec{S}_i \vec{S}_j \rangle = -\frac{1}{4}$. 
This implies an intrinsic tendency of the system to develop low-dimensional
spin fluctuations which can most effectively be realized
by an ordering of orbitals as shown in Fig.~\ref{FIG:ORB}. In this situation
the effective spin interaction is quasi-one-dimensional
along the chains in the c-direction so that quantum spin fluctuations
are enhanced as much as possible and quantum energy is gained
from the bonds along the chain. 
Here $\langle \vec{S}_i \vec{S}_j + \frac{1}{4} \rangle < 0$, so that the effective
orbital (pseudospin) exchange is indeed ferromagnetic, which leads to the
orbital structure shown in Fig.~\ref{FIG:ORB}.
At the same time the cubic symmetry is explicitely
broken, as fluctuations of spin bonds are different in different directions.
This leads to a finite splitting of $e_g$-levels, and therefore an 
orbital gap is generated. One can say that in order to stabilize the ground state, 
orbital order and spin fluctuations support and enhance each 
other  --a situation that is very similar
to Villain's order from disorder phenomena known previously from
frustrated spin systems~\cite{Villain80,SHE82,TSV95}. 

From the technical point of view, it is obvious that a conventional expansion
about the classical N\'eel state would fail to remove the orbital degeneracy: 
only quantum fluctuations can lead to orbital correlations.
This is precisely the reason why in a linear spin-wave approximation 
one does not obtain an orbital gap, and low-energy singularities 
appear~\cite{FEI97,FEI98}. 
The problem was resolved in 
Refs.\cite{KHA97,KHA99}: the singularities vanish once 
quantum spin fluctuations are explicitely taken into account 
in the calculations of the orbiton spectrum.  
These fluctuations generate a finite gap
for single orbital as well as for any composite spin-orbital 
excitation, and in this way the spin fluctuations remove 
the orbital frustration problem. 
The long-range spin-orbital order indicated in Fig.~\ref{FIG:ORB} 
is stable against residual 
interactions because of the orbital gap (of the order of $J/4$), 
and because of the small, but finite, coupling between spin chains.

In general the Kugel-Khomskii model is a very nice example of how an apparently 
three-dimensional system may by itself develop low-dimensio-nal quantum fluctuations. 
We find it quite interesting and amusing that these quantum fluctuations 
do not only coexist with a weak three-dimensional staggered spin moment, but that
they are actually of vital importance to stabilize this long-range order by
generating an orbital gap and intersite orbital correlations. 
The physical origin of this peculiar situation is, of course, the strong 
spatial anisotropy of the $e_g$-orbital wave functions. Because of this
anisotropy it is impossible to optimize all the bonds between Mn$^{3+}$
ions simultaneously; this results in orbital frustration.
The best the system can finally do, is to make specific strong and weak bonds 
in the lattice, whereby it reduces the effective dimensionality of the spin 
system in order to gain quantum energy. 
At the same time tunneling between different orbital configurations is suppressed:
the spin fluctuations produce an energy gap for the rotation of orbitals.

%%%%%%%%%%%%% Orbital models %%%%%%%%%%%%% 

\subsection{Orbital-only models}
\label{Sec:orbitalmodels}
Let us consider for the moment the spins to be frozen in a ferromagnetic configuration,
and ask how the orbitals would behave in this case. 
The model (\ref{HAM2}) then reads 
\begin{equation}
H_{orb} = A \sum_{\langle ij\rangle_\gamma}
T _i^{(\gamma)}T _j^{(\gamma)},
\label{ORB}
\end{equation}
where $T _i^{(\gamma)}$ is given by Eq.~(\ref{PSE}) and 
$A=J/2$. This limit has actually been considered in Ref.\cite{BRI99}
as a model system to describe orbital dynamics in an undoped orbitally
degenerate ferromagnet and for the ferromagnetic
insulating phase of underdoped manganites. In the latter case this mapping is 
not quite satisfying because the ferromagnetic insulator
is in fact stabilized by frozen-in doped holes that strongly affect the orbitals 
in their neighborhood, as was discussed in Section~\ref{Sec:lowdoping}. 
The model (\ref{ORB}) is clearly anisotropic and in that respect underlines
the important difference between orbital excitations and conventional spin dynamics.
Another very closely related model is a so-called ``cubic'' model  
defined as follows:
\begin{equation}
H_{cub}=\frac{A}{4}(\sum_{\langle ij\rangle_a} T_i^x  T_j^x + 
\sum_{\langle ij\rangle_b} T_i^y  T_j^y + 
\sum_{\langle ij\rangle_c} T_i^z  T_j^z),
\label{CUB}
\end{equation}
where along each a,b and c crystallographic axis only one of the 
respective orbital operators $T^x$, $T^y$ and $T^z$ is active. In this
''cubic'' model the bond anisotropy in Hamiltonian~\ref{ORB} is taken
to the extreme. Although the model is not directly applicable to the
manganites, it is simpler than the model in Eq.~(\ref{ORB}), but it still
contains 
the essential bond anisotropy and degeneracy that makes orbital 
models so different from spin Hamiltonians.
This model has been proposed and discussed in the
context of orbital frustrations already in the 1970's~\cite{NOTE1}. 
It is interesting to notice that precisely this
``cubic'' model appears in cubic titanates as a magnetic
anisotropy Hamiltonian~\cite{KHA01}.

We discuss the two models in Eq~(\ref{ORB}),(\ref{CUB}) in 
parallel as they display very similar
peculiarities in the low-energy limit. 
As pseudospins in both models interact
antiferromagnetically along all bonds, a staggered orbital
ordered state is expected to be the ground state of the system. 
In the three-dimensional system at the orbital degeneracy point, however,
linear spin-wave theory leads to a gapless two-dimensional excitation spectrum. 
This results in an apparent instability of the ordered state at any finite 
temperature \cite{BRI99}, an outcome that sounds at least counterintuitive.
Actually, the problem is even more severe: a close inspection shows that
the interaction corrections to the orbiton excitations diverge 
even at zero temperature, manifesting that the linear spin-wave expansion
about a classical staggered orbital, N\'eel-like, state is not adequate in this case.

The origin of these problems was recently clarified in Refs.\cite{KHA01,KHA02}.
By symmetry, there are only a finite number of directions (three equivalent
cubic axes), one of which will be selected by a staggered pseudospin order
parameter. Since this breaks only discrete symmetry, the excitations about
the ordered state must have a gap. 
A linear spin wave theory fails however to give the gap, because
Eqs.(\ref{ORB}),(\ref{CUB}) acquire a rotational symmetry 
in the limit of classical
spins. This results in an infinite degeneracy of classical states,
and an accidental pseudo Goldstone mode appears, which is however
not a symmetry property of the original quantum models (\ref{ORB}),(\ref{CUB}).  
This artificial gapless mode leads to low-energy divergencies that arise
because the coupling constant for the interaction between orbitons does not 
vanish at zero momentum limit, as it would happen for a true Goldstone 
mode. Hence the interaction
effects are non-perturbative. 

At this point the order from disorder mechanism comes again into
play: a particular classical state is selected so that the 
fluctuations about this state maximize the quantum energy gain, 
and a finite gap in the excitation spectra opens, because in the
ground state of the system the rotational invariance is broken. 
An orbital gap $\Delta \simeq A/2$ in the model (\ref{ORB}) has been 
estimated in Ref.~\cite{KHA02,KUB02}, which has to be compared with 
the full orbiton dispersion of $3A$. Similarly, the ``cubic'' model 
with nearest-neighbor interactions shows  long-range orbital order and 
has an excitation gap $\Delta \simeq A$~\cite{KHA01}.
It should be noticed however, that while ground state and
gap issue in models (\ref{ORB}),(\ref{CUB}) are more or less settled, 
further studies are required to fully characterize 
the excitation spectra. Of particular interest
are damping effects, as we expect substantial
incoherent features in the orbiton modes because of the strong interaction  
between low-energy ($\sim \Delta$) orbitons.

What can we learn from the examples above?
The calculation of the excitation spectrum 
in systems with orbital degeneracy 
is somewhat involved even in the half filled, insulating limit.
This is due to the peculiar frustration of 
superexchange interactions, which leads to infrared divergencies
when linear spin wave theory is applied~\cite{FEI97,FEI98,BRI99}.
That such divergencies occur in lowest order approximations 
is a universal feature of $e_g$-orbital models on cubic lattice 
--it reflects the special symmetry properties of $e_g$-orbital pseudospins.
To calculate the excitation spectrum a more careful treatment of quantum 
effects is then required.
 
The main message is that orbital ordering, unlike spin ordering, 
is not accompanied by a simple Goldstone mode. Rather, collective 
orbiton excitations always have
a finite gap and we also expect a substantial incoherent damping 
over all momentum space. This distinct feature 
of the orbitons has to be kept in mind 
for the interpretation of experimental data.
Of course the precise way of how an orbiton gap is generated, depends on the model, 
but an order from disorder scenario seems to be 
a rather common mechanism to resolve the frustration
of classically degenerate orbital configurations.
It is interesting to note that if one takes instead a similar model but with 
long-range interaction, e.g. the classical model with dipole-dipole
interaction, the conclusion about the long-range order might be modified; it was 
shown in~\cite{GLA73} that there is no long-range order of the type shown
in Fig.~\ref{Fig:3dlevels} in the dipole model in two- and three-dimensional systems.

Although this review focuses on $e_g$-orbital dynamics, it is worthwhile 
to mention at this point that orbital frustration is in fact a very general 
property of all orbital degenerate 
cubic perovskite compounds, including those with
three-fold $t_{2g}$ orbital degeneracy. In this case the situation is actually 
even more dramatic because the degeneracy of $t_{2g}$ level is 
larger~\cite{KUG75}, which 
enhances quantum effects~\cite{NOTE2}. 
In addition to the frustration that is present in the $e_g$-system, there also exists
the possibility to form quantum singlets and resonating  valence bonds among $t_{2g}$ 
orbitals (see for the details Refs.~\cite{KHA01,KHA00a}). 
Therefore quantum tunneling between different local orbital configurations may occur.  
The cubic titanate LaTiO$_3$ is an outstanding example:     
it was recently observed~\cite{KEI00} that
orbitals in this Mott insulator remain disordered even at low
temperature, which can be explained by the formation of    
a coherent quantum liquid state~\cite{KHA00a}. The cubic
vanadates are also interesting in this respect, although they are 
different because of the larger spin value. 
In this case the spin-orbital frustration is again resolved by the order from disorder mechanism with the help of low-dimensional orbital fluctuations~\cite{KHA01a}.

%%%%%%%%%%%%%%%%%%%% Doping effects: general %%%%%%%%%%

\subsection{Orbital-charge coupling, orbital polarons}
\label{Sec:polarons}
\begin{figure}
      \epsfxsize=70mm
      \centerline{\epsffile{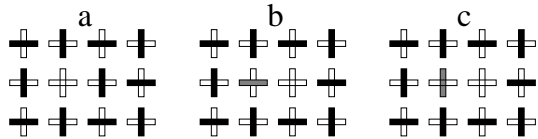}}
\caption{
Schematic representation of a hole in an orbital ordered ground state (a).
The occupied (empty) orbitals are shown as 
filled (empty) rectangles. The hole can move either without disturbing
the orbital order (b), or by creating orbital excitations (c).}
\label{Fig:hop}
%\vspace{-0.5cm}
\end{figure}

We turn now to doping effects in manganites and first discuss the low 
doping regime. 
A doped hole in the Mott insulator strongly interacts with a variety of
low-energy degrees of freedom, which leads to polaronic effects. 
In a pure spin-charge $t-J$ model, as are used for the
cuprates, a  hole breaks the spin bonds, and more importantly, 
frustrates spin order when it hops around. 
In manganites, the presence of orbital
degeneracy brings about new degrees of freedom that control the dynamics of holes.
As is discussed in detail in Ref.\cite{KIL99,BHO00}, there are in general two channels of 
orbital-charge coupling. The first one, that acts via the electron-transfer term of the 
orbital $t-J$ Hamiltonian~\cite{ISH97a}, is very similar to the spin-charge coupling 
mentioned above. 
This interaction of holes with orbital degrees of freedom 
changes the character of the hole motion: the scattering on orbital
excitations leads to a suppression of the coherent quasiparticle
weight $a_{\bar{{q}}}$:~\cite{KIL99}
\begin{equation}
%a_{\bar{\bbox{q}}} = \left\{\begin{array}{l}
a_{\bar{{q}}} = \left\{\begin{array}{l}
\displaystyle
1-\frac{1}{\sqrt{2\pi^2}}\;\left(\frac{t}{J}\right)^{1/2}
%\quad \text{for $J\gg t$},\\[10pt]
\quad for J\gg t,\\
\displaystyle
\sqrt[4]{2\pi^2}\;\left(\frac{J}{t}\right)^{1/4}
\quad for J\ll t.
\end{array}\right.
\end{equation}
In the limit $J/t\rightarrow \infty$ a coherent 
hole motion with $a_{\bar{{q}}}=1$ is recovered. 
In contrast to 
the spin $t-J$ problem in cuprates, the hole mobility  
in this limit is still possible due to the presence a small 
but finite non-diagonal hopping matrix elements which do not 
conserve orbital pseudospin, see Fig~\ref{Fig:hop}.  
Therefore the ``string'' effect~\cite{Bulaevskii68} that occurs
in the spin model,
is less severe in the orbital $t-J$ model.  
In the opposite limit $J/t\rightarrow 0$, however, the holon quasiparticle weight 
is completely lost, which
indicates a strong scattering of holes on orbital fluctuations,
and the excitation spectrum of a doped hole shows only a broad continuum 
in a momentum space. 
On general grounds, it is also expected that 
mobile holes will fill in the orbiton gap, and will eventually
destroy orbital order in a similar way as they melt spin order in
cuprates.

%%%%%%%%% Orbital polarons %%%%%%%%%%%%%%
%
\begin{figure}
\noindent
\centering
\setlength{\unitlength}{0.6\linewidth}
\begin{picture}(1,0.75)
\put(0.355,0.44){\epsfxsize=0.3\unitlength\epsffile{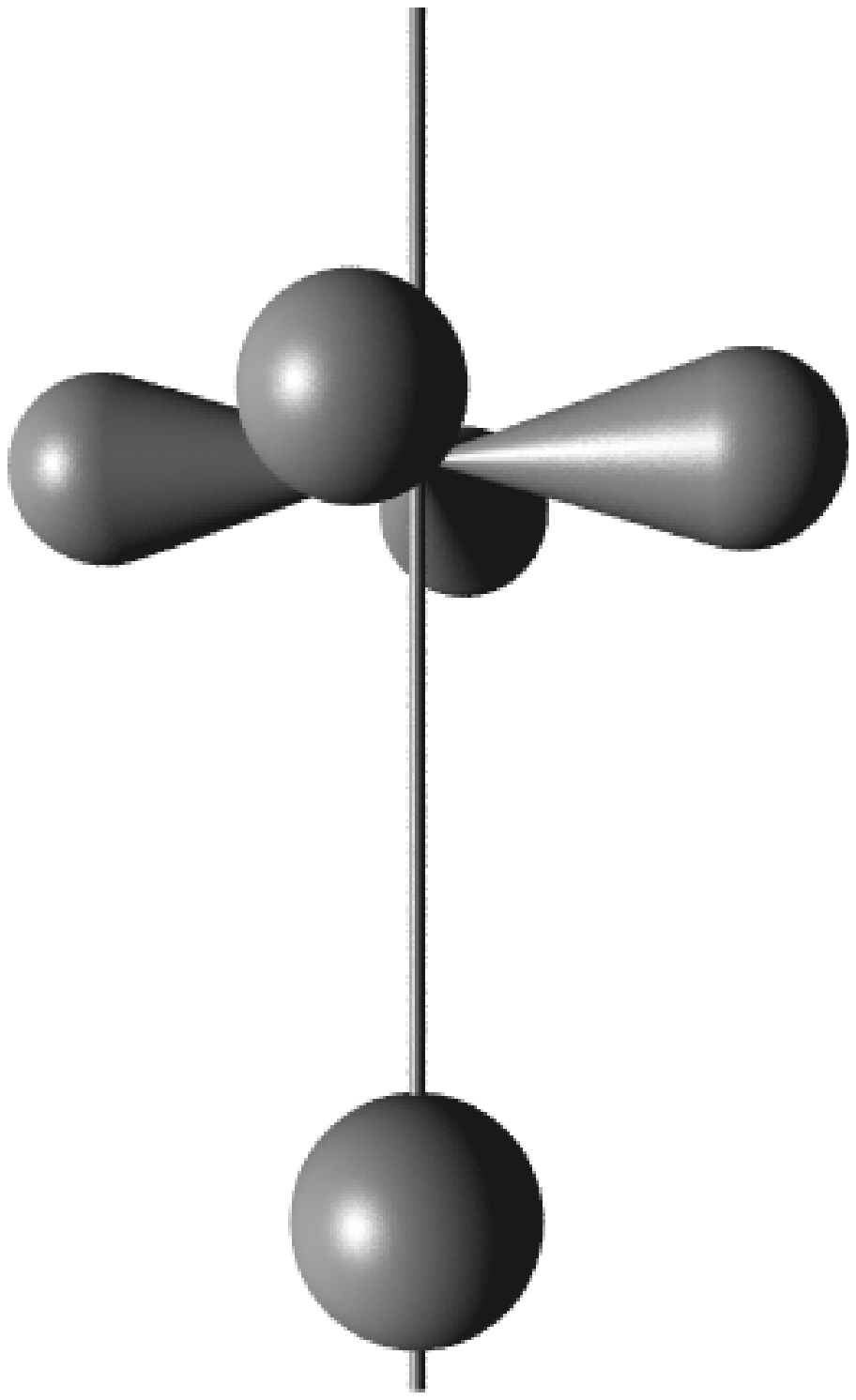}}
\put(0.355,0.01){\epsfxsize=0.3\unitlength\epsffile{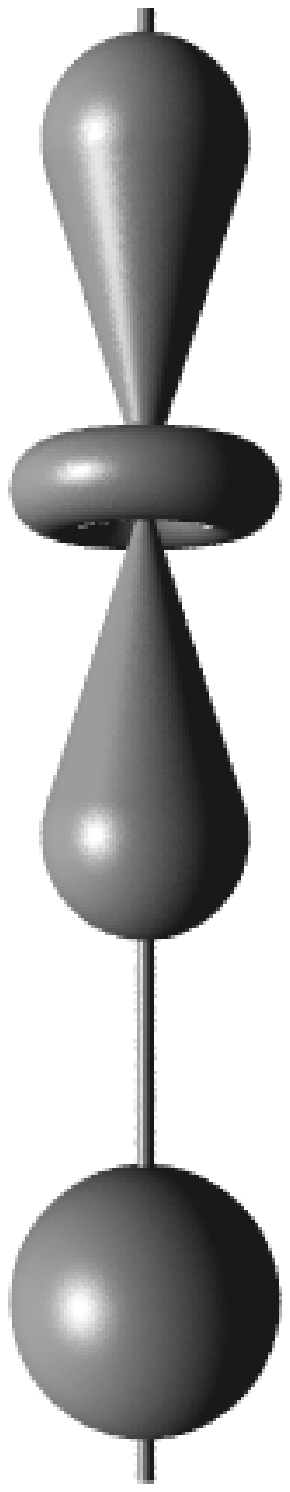}}
\put(0.335,0.44){\fbox{\rule{0.3\unitlength}{0mm}\rule{0mm}{0.3\unitlength}}}
\put(0.335,0.01){\fbox{\rule{0.3\unitlength}{0mm}\rule{0mm}{0.3\unitlength}}}
\put(0,0.15){\epsfxsize=0.85\unitlength\epsffile{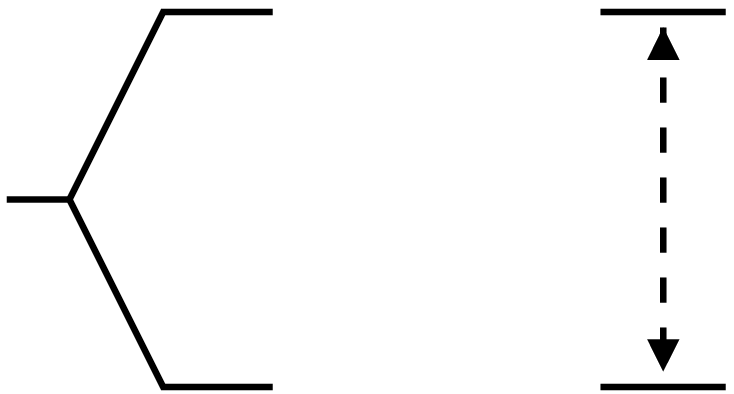}}
\put(0.8,0.36){$\Delta$}
\end{picture}\\[10pt]
\caption{Polarization of $e_g$-levels on sites next to a hole \cite{KIL99}:
Bond stretching phonons and Coulomb interaction induce a splitting of energy
$\Delta=\Delta^{\rm{ph}}+\Delta^{\rm{ch}}$. Here the sphere indicates the location
of a hole (Mn$^{4+}$-ion) adjacent to a Mn$^{3+}$-ion.}
\label{FIG:SPL}
\end{figure}

It turns out, however, that the spatial anisotropy of $e_g$-orbitals
provides yet another, very important channel of 
orbital-charge coupling which
is very specific to the case of manganites. 
Notice that, different from the cuprates, holes remain 
localized up to rather high doping levels, and at the 
same time induce an isotropic ferromagnetic state. The complete breakdown 
of metalicity at hole concentrations below $x_{\rm{crit}} \approx 
0.15$ - $0.2$ occurs despite the fact that ferromagnetism is
fully sustained, and is sometimes even stronger, 
in this regime~\cite{endoh,Uhlenbruck99}, which seems very surprising 
from the point of view a standard double-exchange picture.
To explain this puzzle, the concept of orbital polarons was introduced
in Ref. \cite{KIL99} and used for the low-doped manganites in~\cite{mizokawa,mizokawa2}.
We discussed the orbital polaron concept in Section~\ref{Sec:lowdoping} for the purpose
of describing possible orbital and charge ordering at fractional dopant concentrations, 
in the low doping regime. Here we treat this topic on a somewhat more theoretical level,
also because orbital polaron formation can be viewed as a precursor for the orbital
liquid state --these are competing states-- 
that can occur in optimally doped manganites, which we discuss in 
the following Section.

The important point is that in an orbitally
degenerate Mott-Hubbard system there also exists a specific coupling
between holes and orbitals that stems from the polarization of 
$e_g$-orbitals in the neighborhood of a hole. 
In this way the cubic symmetry at manganese ions close to the hole
is lifted, see Fig.~\ref{FIG:SPL}).
The displacement of oxygen ions, the Coulomb force exerted by the positively 
charged hole and hybridization of electrons 
cause a splitting of orbital levels. This splitting is 
comparable in magnitude to the kinetic energy of holes so that the orbital-hole 
binding energy can be large enough for holes and surrounding orbitals to 
form a bound state. 
\begin{figure}
\noindent
\centering
\epsfxsize=0.6\linewidth
\epsffile{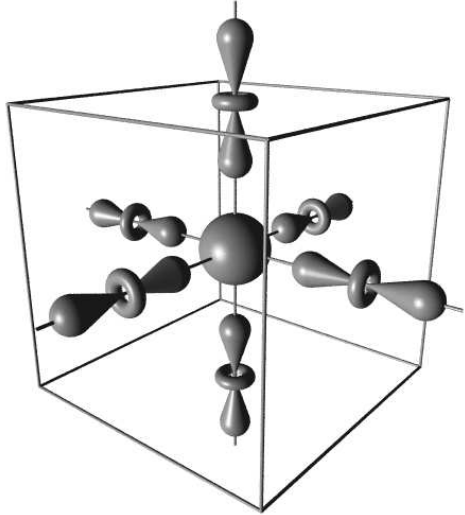}\\[6pt]
\caption{Orbital polaron in the strong-coupling limit: Six $e_g$-
states point towards a central hole \cite{KIL99}.}
\label{FIG:OPO}
\end{figure}
 For a given bond along the $z$ direction the interaction reads as 
$H^z = -\frac{1}{2}\Delta \;n^h_i\sigma^z_j$. 
This is precisely the new orbital-charge coupling channel which is of course 
absent in pure spin-charge models.
The splitting of $e_g$-levels effects all six sites surrounding a hole, and
the analogous expressions for $x$ and $y$ directions 
can easily be derived by a rotation of the interaction Hamiltonian in orbital space. 
The complete orbital-charge coupling Hamiltonian for the cubic system is then
\begin{equation}
H_{\rm{ch-orb}} = -\Delta \sum_{\langle ij \rangle_{\gamma}}
n_i^h T_j^{(\gamma)},
\label{HPO}
\end{equation}
with orbital pseudospin operators given by Eq.~(\ref{PSE}).
The interactions in Hamiltonian (\ref{HPO}) describe the tendency of the system
to form orbital polarons.
For low enough hole concentrations the polaron consist of a bound state between
a central hole with surrounding $e_g$-orbitals pointing towards the hole
as is shown in Fig.~\ref{FIG:OPO}.
 
The structure of the orbital polaron yields a large amplitude of virtual excursions of
$e_g$-electrons onto the empty site. Thus, besides minimizing the 
interaction energy of Hamiltonian (\ref{HPO}), the polaron also allows for a lower
kinetic energy. We note that these virtual hopping processes locally
enhance the magnetic moments of core and $e_g$-spins via the 
double-exchange mechanism in all three directions and provide 
a large effective spin of the orbital polaron. 
This naturally explains the development of 
ferromagnetic clusters experimentally observed at temperatures above 
$T_C$ at low doping levels~\cite{TER97}. At finite hole densities these clusters
start to interact, thereby inducing a global ferromagnetic state.
Clearly, the orbital state of such a ferromagnetic insulator must be
very complex due to the presence of frozen-in orbital polarons.
The relevant model must contain the interactions given by both
Eqs. (\ref{HPO}) and (\ref{ORB}) in a first step, and further
be complemented by, at least, the Coulomb forces between holes.  
It was suggested~\cite{KIL99} that a such state may have 
orbital/Jahn-Teller glass features, which reduces the long-range
component of static Jahn-Teller distortions. The orbital excitations
are expected therefore to have a large broadening in a momentum space.
These issues clearly deserve more theoretical work, particularly in 
the perspective of recent experimental results on the observation of
orbital excitations in manganites~\cite{saitoh}, which hopefully can
be extented to the lightly doped ferromagnetic insulator regime.
    
The formation of an orbital polarons competes with kinetic energy of holes 
that tends to delocalize the charge carriers, but also competes with
the fluctuation rate $\propto xt$ of orbitals: the faster the orbitals
fluctuate, the less favorable it is to form a bound state in which orbitals
have to give up part of their fluctuation energy. In other words, the binding
energy of the polaron decreases at higher doping levels. 
When this orbital polaron
picture is combined with that of conventional lattice polarons~\cite{MIL96,ROD96}, 
the transition from ferromagnetic metal to ferromagnetic 
insulator can be well explained~\cite{KIL99}.
 
Once the orbital polaron is formed it will be trapped by even a weak disorder.
Near commensurate fillings, say about 1/8, they may form 
a polaron superlattice, optimizing simultaneously the charge 
and orbital configuration energies, as we discussed 
in section~\ref{Sec:lowdoping}. Commensuration through phase separation
--which is certainly relevant in manganites \cite{MOR99}-- is also possible,
but we will not review this complex issue here.  

%Needless to say, the orbital polaron of the sort in Fig.~\ref{FIG:OPO}
%can not form at large dopings for a simply space counting arguments. Yet the
%orbital-hole interaction Eq.~(\ref{HPO}) is at the work yielding a dynamical
%splitting of the $e_g$-levels thus enhancing orbital disorder. 
 
%%%%%%%%% Optical Conductivity %%%%%%%%%%%%%%

\subsection{Orbital liquids, anomalous transport}
\label{Sec:liquid}

An almost universal feature of manganites is that at higher doping concentrations 
(around $x \approx 1/3$) a ferromagnetic metallic state emerges. 
In this section we discuss the orbital state and orbital fluctuations 
in this regime.
The appearance of a ferromagnetic metallic state is 
explained in a framework of double exchange 
physics~\cite{degennes,ZEN51,AND55,KUB72}.
But if one takes the double exchange model as the starting point to explain
the properties of the metallic state, it is quite surprising that in experiment
one finds that the $e_g$-electrons do not behave as conventional
spin-polarized carriers in a uniform ferromagnetic phase at all. 
For such carriers one expects that optical spectral weight is accumulated into 
a low-frequency Drude peak, but the weight of optical
spectra in manganites robustly extends up to $\sim 1$eV, even at very 
low temperatures~\cite{OKI95}.
 
This experimental finding is remarkable for at least two reasons.
First, the energy scale extends to the electron-volt range, which 
rules out a purely phononic origin of the incoherence; and second, the
incoherent spectral weight is very large in magnitude, even at low temperatures. 

Other experimental studies show that collective as well as local lattice 
distortions are absent in metallic manganites at low temperature, 
which may be an indication that orbital fluctuations are strong. 
Based upon this observation, 
several authors~\cite{ishihara97,SHI97} attributed the incoherent structure of the
optical conductivity to the orbital degrees of freedom. While the 
study~\cite{SHI97} is based on a simple band picture, a more elaborate treatment
of both orbital degeneracy and on-site correlations was suggested in
Ref.~\cite{ishihara97}. In this work the notion of an orbital liquid, 
which describes a quantum disordered state of $e_g$-quadrupole moments in 
metallic manganites, was introduced.
Such quantum disorder of orbitals is caused by the motion of holes which mixes up 
dynamically all possible orbital configurations. 

At first glance this idea seems to be very similar to the concept of the moving
holes that cause spin disorder in cuprates. 
There are two differences, however.
The first one is that in a ferromagnetic system the orbital superexchange interactions 
are rather frustrated from the very beginning and this actually favors the orbital 
liquid, as we discussed previously. 
The effective two-dimensionality of the pseudospin fluctuations is, in fact,
emphasized in~\cite{ishihara97} as an important disorder mechanism. 
While indeed enhancing quantum effects, this effect alone would 
still not be sufficient to destroy the pseudospin order, 
as one can see from the behavior of orbital models discussed above.
Rather, it acts cooperatively with the frustrations induced by charge motion.  
The second, and most important, difference from cuprates is the strong tendency
for holes in manganites to localize and form orbital/lattice polarons. 
It is the restricted charge mobility that confines the orbital liquid metallic 
state of manganites to a rather small region of the phase diagram.

The optical 
conductivity in the metallic phase of 
manganites has been calculated in Ref.~\cite{KIL98}, taking the
idea of an orbital liquid as a starting point.  
The strongly correlated nature of the $e_g$-electrons in the metallic system
can conveniently be accounted for by employing a slave-boson 
representation of electron operators:
\[
c^{\dagger}_{i\alpha} = f^{\dagger}_{i\alpha}b_i.
\]
Here the orbital pseudospin is carried by fermionic orbitons 
$f_{i\alpha}$, and charge by bosonic holons $b_i$. In comparison with
other treatments of the orbital $t-J$ model~\cite{ishihara97,ISH97a}, 
this representation is most adapted to describe the 
correlated Fermi-liquid state of manganites 
since it naturally captures both the coherent and incoherent features
of excitations. The small Drude peak as well as a broad optical absorption
spectrum, that extends up to the bare bandwidth, are well reproduced by 
these calculations. The physical picture is that the charge carriers
scatter strongly on the dynamical disorder caused by the fluctuations
of orbital bonds that are due to the correlated rotation of orbitals.  
The fact that the anomalous transport properties
in the ferromagnetic metallic phase can be described consistently, 
supports the validity of the orbital-liquid framework.

%%%%%%%%%%%% Magnons %%%%%%%%%%%%%%%%%%%%%%%%

\subsection{Fluctuating bonds, magnon anomalies} 
\label{Sec:magnon}
\begin{figure}
\centering
\setlength{\unitlength}{0.18\linewidth}
\rule{0.15\unitlength}{0\unitlength}
\begin{picture}(1,1.92)
\put(0,0){\epsfxsize=1.0\unitlength\epsffile{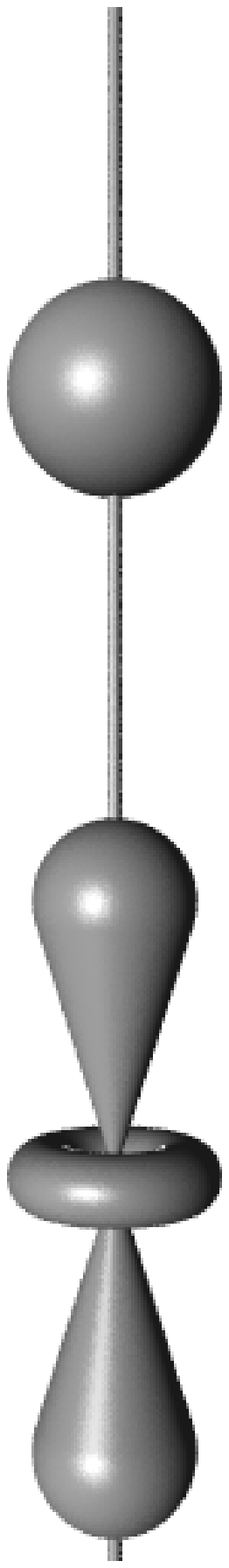}}
\put(-0.15,0.6){\epsfxsize=0.4\unitlength\epsffile{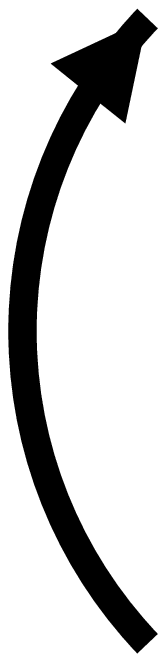}}
\put(0.1,-0.35){$J_{DE}^c \propto t$}
\end{picture}
\rule{0.4\unitlength}{0\unitlength}
\rule{0.45\unitlength}{0\unitlength}
\begin{picture}(1,2.14)
\put(0,0){\epsfxsize=1.0\unitlength\epsffile{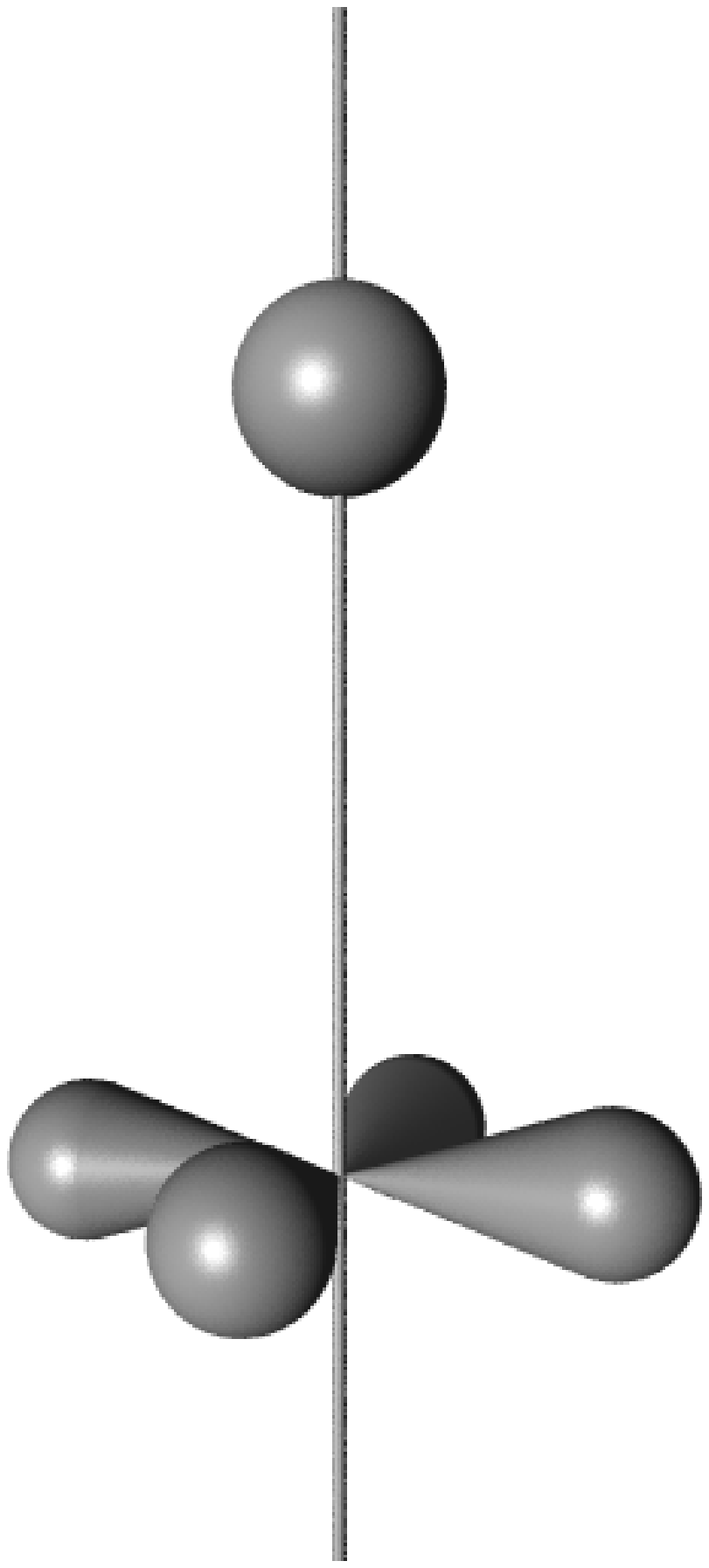}}
\put(-0.45,0.6){\epsfxsize=0.4\unitlength\epsffile{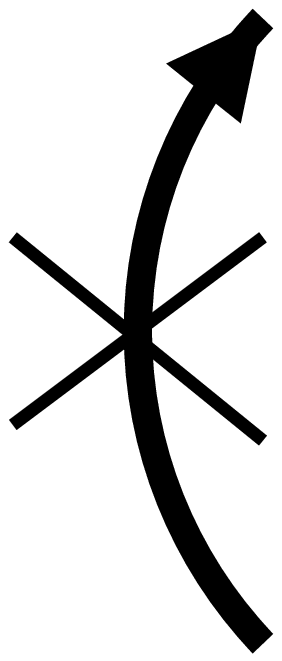}}
\put(0.1,-0.35){$J_{DE}^c = 0$}
\end{picture}
\rule{0.15\unitlength}{0\unitlength}\\[0.65\unitlength]
\caption{The $e_g$-electron transfer amplitude, which controls the
double-exchange interaction $J_{\rm{DE}}$, strongly depends on
the orbital orientation: along 
the $z$ direction, e.g., $d_{3z^2-r^2}$ 
electrons (left) can hop into empty sites denoted by a sphere, 
while the transfer of $d_{x^2-y^2}$ electrons (right) 
is forbidden \cite{KHA00}.}
\label{FIG:MBD}
\end{figure}

Finally, we discuss one more interesting manifestation
of the orbital fluctuations in ferromagnetic metallic manganites.
According to the conventional theory of double 
exchange, the spin dynamics of the ferromagnetic state 
%that evolves at temperatures below the Curie 
%temperature $T_C$ 
is expected to be of nearest-neighbor Heisenberg type with
a simple cosine-like magnon dispersion~\cite{FUR96}. 
This picture seems indeed to be reasonably accurate for manganese oxides
with high Curie temperature $T_C$, i.e., for compounds with a
ferromagnetic metallic phase that is sustained up to rather high 
temperatures~\cite{PER96}.
Recent experimental studies on compounds with low values
of $T_C$ indicate, however, marked deviations from this canonical behavior. 
Quite prominent in this respect are measurements of the spin 
dynamics of the ferromagnetic manganese oxide 
Pr$_{0.63}$Sr$_{0.37}$MnO$_3$~\cite{HWA98}. 
While the spin excitation spectra exhibit conventional Heisenberg behavior 
at small momenta, the dispersion of magnetic excitations (magnons) shows a
curious softening at the boundary of the Brillouin zone in the [1,0,0] and
[1,1,0] direction but {\it not} in the [1,1,1] direction. 
The origin of this unusual softening has been attributed  
in Ref.\cite{KHA00} to the low-energy orbital fluctuations 
present in a ``narrow-band'' manganites.  
Let us briefly explain the basic idea. 
The strength of the ferromagnetic interaction at a given bond strongly depends 
on the orbital
state of $e_g$-electrons (see Fig.~\ref{FIG:MBD}). 
Along the $z$ direction, for instance, only electrons in $d_{3z^2-r^2}$ orbitals 
can hop between sites and hence can participate in the double-exchange processes,
but the transfer of $d_{x^2-y^2}$ electrons is blocked due to the vanishing overlap
with O-$2p$ orbitals located in-between two neighboring Mn sites. 
Temporal fluctuations of $e_g$-orbitals may thus
modulate the magnetic exchange bonds, thereby
renormalizing the magnon dispersion. Actually, such an effect is a quite
general property of orbitally degenerate systems; the same kind of
zone-boundary softening has been predicted in the insulating Kugel-Khomskii
model as well~\cite{KHA97,KHA99,stekelen}.

Short-wavelength magnons are
most sensitive to these local fluctuations and are affected most strongly. 
Quantitatively the modulation of exchange bonds is controlled by the 
characteristic time scale of orbital fluctuations: if the typical 
frequency of orbital fluctuations is higher than the timescale for  
spin fluctuations, the magnon spectrum remains mostly unrenormalized.
In this case the orbital state effectively enters the spin dynamics only 
as a time average, which restores the cubic symmetry of exchange bonds.
On the other hand, if orbitals fluctuate slower than spins, then the 
renormalization of the magnon spectrum is most pronounced 
and the anisotropy imposed upon the magnetic exchange bonds by the orbital 
degree of freedom comes into play. 
Also a damping of magnons is then expected to occur,
which may explain the large broadening of magnons at the zone boundary.  

We note that the magnons in $[1,1,1]$ direction are sensitive to 
all three spatial directions of the exchange bonds; their dispersion
therefore remains unaffected by the local dynamical symmetry breaking 
induced by low-dimensional orbital correlations. This leads
to a strong anisotropy of magnon renormalization effects in a momentum
space: the $[1,0,0]$, $[1,1,0]$ directions are mostly
affected, which is actually in agreement with experiment~\cite{HWA98}.
The momentum and energy dependence of magnon anomalies are in general very 
sensitive to the character of local orbital correlations, and may therefore 
change with the evolution of underlying orbital states. In other words, if the 
[0,0,1] magnon mode softens to zero at the zone boundary, 
than this would signal the transition from ferromagnetic 
to the A-type antiferromagnetic spin ordering.
 
The unusual magnon dispersion experimentally observed in low-$T_C$
manganites can hence be understood as a precursor effect of
orbital/lattice ordering. While the
softening of magnons at the zone boundary is responsible
for reducing the value of $T_C$, the small-momentum spin 
dynamics that enters the spin-wave stiffness $D$ remains 
essentially unaffected. This explains the enhancement of
the ratio $D/T_C$ observed in low-$T_C$ compounds~\cite{FER98}.

%%%%%%%%%%%%%Conclusion %%%%%%%%%%%%%%

\subsection{Summarizing remarks}
\label{Sec:summary}
We can state in general that quantum fluctuations of orbital degrees of freedom 
in manganites are important and have to be taken into account in the explanation of 
various properties of these systems. 
Despite (or even due to) the absence of the rotational symmetry in the orbital sector, 
profound non-classical behavior of orbitals can be caused by the 
frustrating nature of interactions in the orbital sector and by the strong 
coupling of orbitals to doped holes.  

We did not discuss the consequences of electron-phonon coupling on the
orbital liquid state, so the question arises how phonons would change the 
physics we discussed in this section. 
Breathing phonons, which are certainly important for a charge localization, 
are actually implemented in the above picture through the interaction 
in Eq.~(\ref{HPO}).   
In general we expect that the coupling of the orbitals to Jahn-Teller phonons 
cooperates with the superexchange process in establishing orbital 
order and local orbital correlations. However, the fact that 
Jahn-Teller phonon frequencies may fall into the region of electronic orbiton 
energies makes the problem very difficult to analyze, particularly at the
proximity to the insulator-metal transition
(which is, simultaneously, an orbital solid-liquid transitions) 
in the phase diagram of manganites.
%This problem has therefore been left out of the present discussion 
%of quantum dynamics of orbitals.

\section{Orbital order for $x \ge 1/2$}
\label{Sec:x.ge.half}

In the phase diagram of the manganites (Fig.~\ref{Fig:phase}) there is an
apparent asymmetry between the case $x<1/2$ and the case $x>1/2$. At lower doping levels
ferromagnetic states dominate, one of which is metallic, and for high doping
concentrations antiferromagnetic insulating phases dominate. Exactly at 
$x=1/2$ for most manganites a rather remarkable antiferromagnetic
(CE-type) charge-ordered insulating state is usually realized. In section~\ref{Sec:x.gt.half} 
we discuss the properties of manganites at high doping levels and why they 
are so different from the systems with less doped carriers. 
In the next section we describe the situation at half filling and
discuss the actual mechanism that leads to the CE-type 
(charge, orbital and magnetic) ordering in half-doped manganites.

\subsection{Half-doped manganites}
\label{Sec:x.eq.half}

\begin{figure}
      \epsfysize=65mm
      \centerline{\epsffile{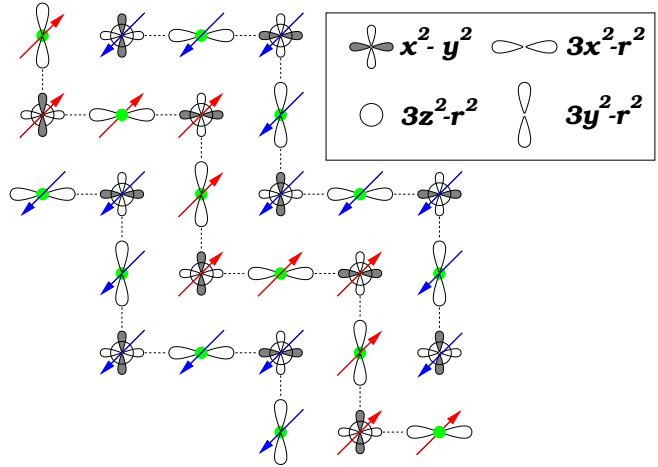}}
\caption{Detailed view of the CE-phase in the x-y plane. We choose our basis orbitals such that 
the gray lobes of the shown orbitals have a negative sign.
The dots at the bridge-sites represent a charge-surplus.}
\label{Fig:ce_phase2}
\end{figure}

The half-doped manganites with $x=1/2$ are very particular. Magnetically
these systems form ferromagnetic zig-zag chains that are
coupled antiferromagnetically (see Fig.~\ref{Fig:ce_phase} and Fig.~\ref{Fig:ce_phase2}) 
at low temperatures, the so-called magnetic CE-phase~\cite{wollan}.
The ground state is, moreover, an orbitally ordered and charge-ordered insulator. This
behavior is generic and is experimentally observed in 
Nd$_{1/2}$Sr$_{1/2}$MnO$_3$~\cite{Kuwahara95,Kawano97},
Pr$_{1/2}$Ca$_{1/2}$MnO$_3$~\cite{Tomioka96}, 
La$_{1/2}$Ca$_{1/2}$MnO$_3$ ~\cite{Urushibara95,Mori98},  
Nd$_{1/2}$Ca$_{1/2}$MnO$_3$~\cite{Kimura_prep}
and in the half-doped layered manganite 
La$_{1/2}$Sr$_{3/2}$MnO$_4$~\cite{Moritomo95}.
The insulating charge-ordered state can be transformed into a metallic FM state by
application of an external magnetic field, a transition that is accompanied by a change
in resistivity of several orders of magnitude~\cite{Kuwahara95,Tokunaga98}.

The extraordinary properties of the half-doped manganites are due to the commensurability
of the doping level. The disorder caused by orbital fluctuations can be quenched by magnetic, 
lattice and charge instabilities that are most pronounced at this commensurate filling.
The combining effect of these three instabilities leads to an effectively
one-dimensional insulating state, 
to the charge ordering and to the unusual magnetic ordering in these systems.

\begin{figure}
\noindent
\centering
\setlength{\unitlength}{0.7\linewidth}
\begin{picture}(1,0.5)
\put(-0.2,0.2){\epsfxsize=0.6\unitlength\epsffile{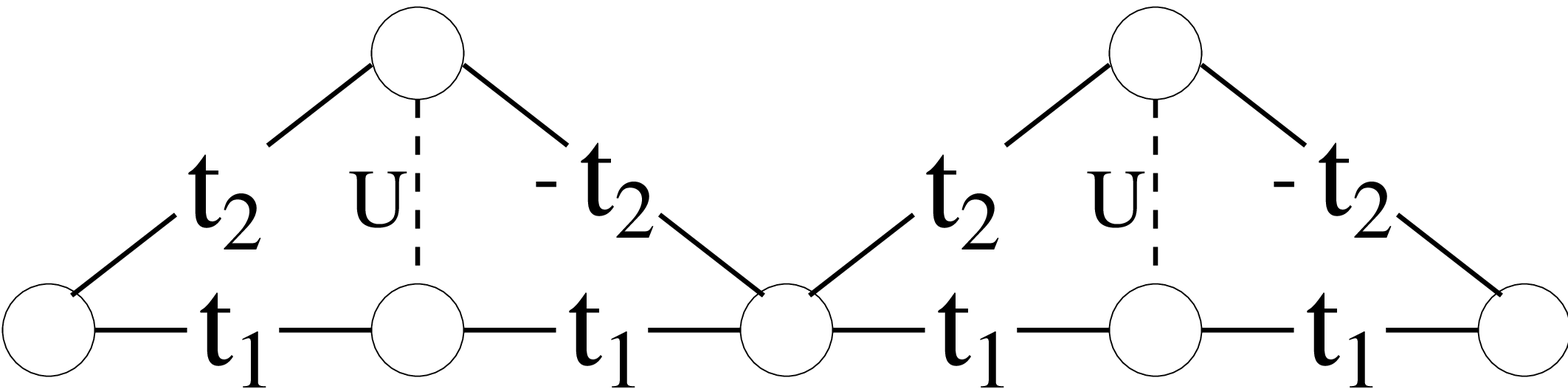}}
\put(0.5,0){\epsfxsize=0.6\unitlength\epsffile{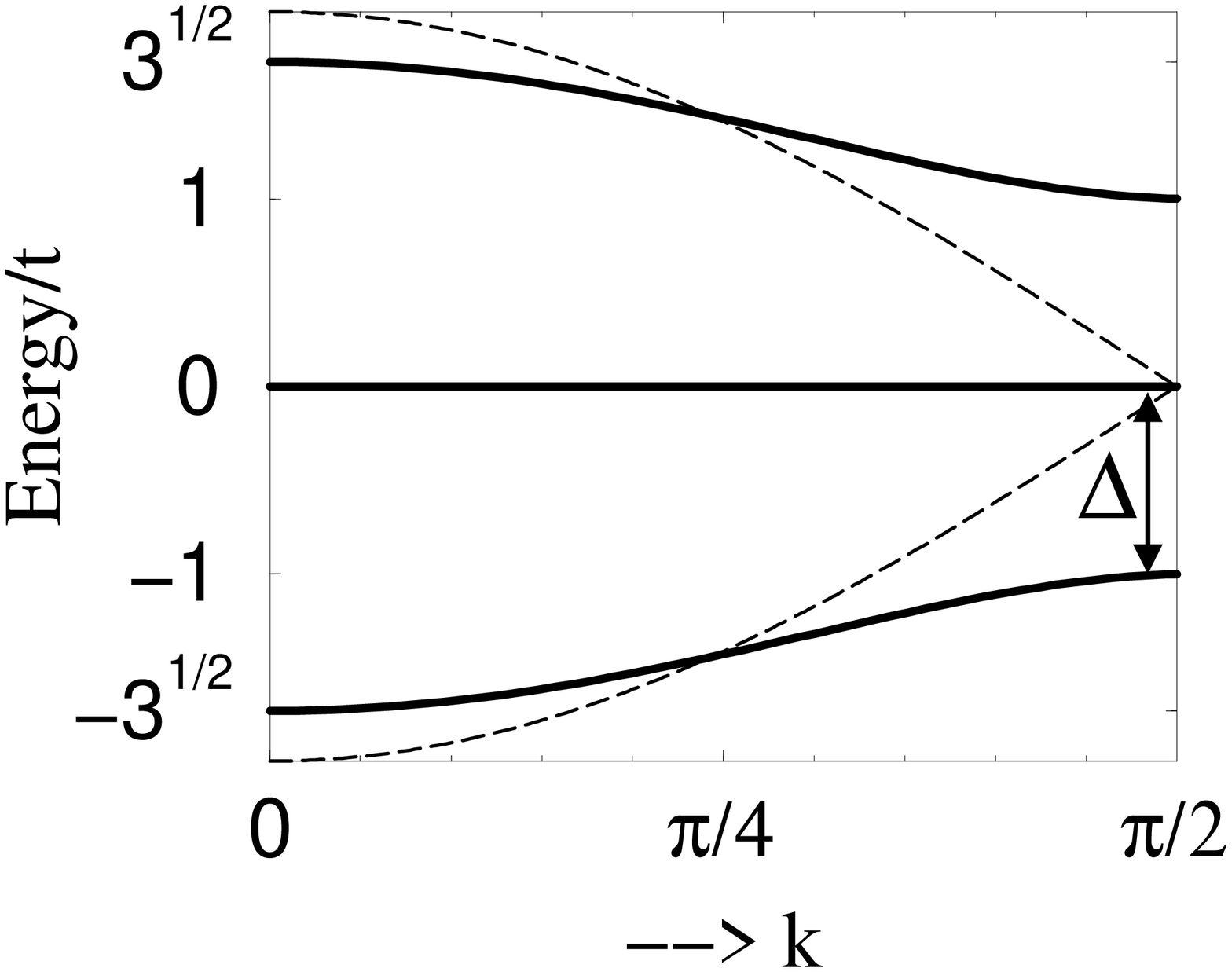}}
\end{picture}\\[10pt]
\caption{Left: Topology of the interactions in a zig-zag chain, where $t_1=t/2$, $t_2= t \sqrt{3} /2$, 
and $U$ is the Coulomb interaction between electrons on the same site.
Right: electron dispersion in the zig-zag chain of the CE-phase for $U=0$ (solid lines)
and electron dispersion 
in a straight chain, as in C-phase (dashed lines).}
\label{Fig:hopCE}
\end{figure}

In the double-exchange framework electrons can only hop between sites with FM aligned
core-spins so that in the CE-phase only hopping processes within the zig-zag chains are
possible, rendering the system effectively one-dimensional for low-energy charge 
fluctuations~\cite{soloviev,vandenbrink}. 
The unit cell of the quasi-1D system contains two Mn atoms, one situated at a corner
site and one situated at a bridge-site, between two corners. As there are two orbitals 
per site, the cell has in total four different orbitals, and at half doping we have on average 
one electron per unit cell. 
The topology of the electron hopping integrals between orbitals is  shown in Fig.~\ref{Fig:hopCE}.
The important observation is that an electron that hops from one bridge-site to another
bridge-site via a $|x^2-y^2 \rangle $ corner-orbital obtains a phase-factor $-1$, while if the
hopping takes place via a $| 3z^2-r^2 \rangle $ corner-orbital, the phase-factor is $+1$. 
This phase factor can be viewed as an effective dimerization that splits the four bands.
These bands are shown in Fig.~\ref{Fig:hopCE}, where
there are two bands with energy $\epsilon_{\pm}= \pm t\sqrt{2 + \cos k}$, where
$k$ is the wave vector, and degenerate two dispersionless bands at zero energy.
At $x=\frac{1}{2}$ the $\epsilon_{-}$ band is fully occupied, and all other bands are empty.
The system is insulating as the occupied and empty bands are split by a gap $\Delta=t$
and the minority bands with spin opposite to the ferromagnetic orientation of the chain 
(not shown) are about $J_H$ higher in energy ($J_H$ is the Hund's rule exchange energy). 
The gap is
very robust as it is a consequence of the staggered phase-factor that itself is
fully determined by the topology of the system.
In principle other quasi 1D magnetic states, e.g. antiferromagnetically coupled
ferromagnetic straight chains, are also possible; they would correspond to a state
with all orbitals of the Mn-sites e.g. of the  $3x^2-r^2$ type, instead of alternating
$3x^2-r^2$ and $3y^2-r^2$ orbitals.
But the presence of a gap
stabilizes the CE-phase. This mechanism is equivalent to the situation in the 
lattice-Peierls problem, where the opening of a gap stabilizes a ground state with a lattice 
deformation. An extra reason for the stability of this phase maybe be the elastic
interactions~\cite{khomskii3,khomskii4}. We see the importance of the commensurability: for $x>1/2$
the excess holes deplete the valence band, reducing the effect of the energy gain due
to the gap. One expects that in this case other magnetic phases, that do not support 
such a gap, compete strongly with the CE-phase.
If on the other hand $x<1/2$, the excess electrons will enter in non-dispersive bands and 
would not contribute to the kinetic energy of the system and would 
thereby effectively destabilize 
the CE-phase. In fact it was argued that in this case the system is 
phase separated~\cite{vandenbrink,Babushkina99,Kagan01}. This agrees with the fact that the
CE-phase survives in Pr$_{1-x}$Ca$_x$MnO$_3$ to $x=0.3$.

But why should the system organize itself into a quasi one-dimensional state in the first 
place? It is rather obvious that in a three-dimensional state, where there is much more
freedom for an electron to move around, the kinetic energy would be lower. The answer is that
in the manganites there are two competing magnetic interactions: the double exchange, 
favoring ferromagnetism, and the superexchange, that is a driving force for 
antiferromagnetism. The system can gain energy from both interactions
simultaneously when in some direction antiferro. bonds are formed, and in other directions
--ferro. bonds. Electrons can then only propagate along
ferromagnetic bonds because of the double exchange. 
This mechanism is especially effective for systems with $e_g$-orbitals
and is in fact quite common in the highly doped manganites --we will 
discuss this in the next section.
The situation reminds us of the order from disorder scenario in spin-orbital
models discussed in section~\ref{Sec:models}: by forming different kinds of bonds 
--which is quite natural for spatially strongly anisotropic $e_g$-orbitals-- 
the system gains energy.

In the CE-phase only the elongated orbitals at the bridge site (the $3x^2-r^2$ and
$3y^2-r^2$ ones)
are occupied 
(see Fig.~\ref{Fig:ce_phase2}). 
The reason is that because of the symmetry the orthogonal, planar orbitals on the
bridge sites do
not have any overlap with orbitals at the corner sites and in first approximation 
do not interact with the rest of the system. On the corner sites both orbitals are in
principle partially occupied. The different orbital occupation on bridge and corner
site --the orbital order-- causes a lattice deformation via the Jahn-Teller coupling,
thereby lowering the energy of the system still further. 

Let us now discuss the consequences of electron-electron interactions at half filling. 
In principle one
expect that short range electronic density-density interactions, that lead to the
Mott insulator at zero doping, have less impact on physical
properties when doping is increased, simply because the density of electrons becomes
smaller and the electrons do not encounter each other very often. 

As we pointed out above, the CE-phase is orbitally ordered, but charge is homogeneously 
distributed between corner and bridge sites if Coulomb interactions are neglected.
It is well known that longer-range Coulomb interactions (for instance a nearest neighbor 
interaction) can cause charge ordering, especially at commensurate filling.
A surprising observation, however, 
is that the experimentally observed charge order can be directly obtained
from the degenerate double-exchange model when only the Coulomb interaction 
(the Hubbard $U$) between electrons in different orbitals, but {\it on the same site}
is included~\cite{vandenbrink}. 
This can be understood from the fact that in the band picture on the corner sites
both orthogonal orbitals are partially occupied, but on the bridge sites only one orbital
is partially filled. The on-site Coulomb interaction acts therefore differently on the
corner and bridge sites: charge is pushed away from the effectively correlated
corner sites to the effectively uncorrelated bridge-sites.

The on-site Hubbard $U$ thus leads to intersite charge disproportionation. Long-range
Coulomb interactions will of course strengthen this charge ordering, and the Jahn-Teller
distortion, which lowers the on-site energy of the bridge orbital with respect to
the corners, also contributes to the ordering. Whereas the effect of $U$ is to increase
the ground state energy with respect to the C-phase~\cite{commentCE} 
(which is made up of straight ferromagnetic 
chains), the longer-range Coulomb interactions will stabilize the CE-phase. This indicates 
that there are several competing kinetic, potential and lattice contributions to the total 
energy and that one needs to consider these in detail to determine the actual phase 
diagram~\cite{replyCE}.

\subsection{Overdoped manganites}
\label{Sec:x.gt.half}
Now we qualitatively discuss the role of orbital degrees of
freedom in the overdoped regime, $x>0.5$. The main question is why 
 the conventional double exchange, apparently responsible
for the formation of the ferromagnetic metallic state for $x\sim0.3$
-- $0.4$, does not lead to such a state in this case.

One reason may be the following. Usually we ascribe ferromagnetism in
doped systems to a tendency to gain kinetic energy by maximal
delocalization of doped charge carriers. These carriers are holes in
lightly doped manganites  $x < 1$  , and electrons when we start
e.g.\ from CaMnO$_3$ and substitute part of Ca by La or other rare
earths, which corresponds to $x<1$ in La$_{1-x}$Ca$_x$MnO$_3$.

There exist an important difference between these two cases, however.
When we dope LaMnO$_3$, the orbital degeneracy in the ground state is
already lifted by orbital ordering, and in a first approximation
we can consider the motion of doped holes in a nondegenerate band.
Then all the standard treatment, e.g. that of de~Gennes~\cite{degennes},
applies, and we get the FM state. However, when we start from the
cubic CaMnO$_3$, we put extra electrons into empty
{\it degenerate} $e_g$-levels, which form degenerate bands.
Therefore we have to generalize the conventional double-exchange
model to the case of degenerate bands. This was done
in~\cite{vandenbrink2}, and the outcome is the following: at
relatively low electron concentration ($x\simeq1$) the anisotropic
magnetic structures --C-type (chain-like) or A-type (ferromagnetic
planes stacked antiferromagnetically)-- are stabilized, and only
close to $x\sim0.5$ we reach the ferromagnetic state. The C-phase
occupies larger part of the phase  space. The resulting theoretical
phase diagram~\cite{vandenbrink2} is in surprisingly good agreement
with the properties of Nd$_{1-x}$Sr$_x$ MnO$_3$~\cite{kuwahara} in
which there exist the A-type ``bad metal'' phase for $0.52<x<0.65$
and C-phase for $x>0.65$.

A simple qualitative explanation of this tendency is the
following. When we start from CaMnO$_3$ with Mn${^{4+}}$
($t^3_{2g}$)-ions and dope it by electrons, we put electrons into
$e_g$-bands. The maximum energy we can gain is to put these electrons
at the bottom of corresponding bands, so that one has to make these
bands as broad as possible. But due to a specific character of the
overlap of different $e_g$-orbitals in different directions, the
bottom of the bands coincides for different types of orbitals: one
can easily check that if we make all the orbitals e.g.\ $3z^2-r^2$, the
energy $\epsilon(k)$ at the $\Gamma$-point $k=0$ will be the same as
for the bands made of ($x^2-y^2$)-orbitals. (Actually it is a
consequence of the degeneracy of $e_g$-orbitals in cubic crystals:
the symmetry at the $\Gamma$-point should coincide with the point
symmetry of local orbitals, i.e.\ at $k=0$ the energies of the
$3z^2-r^2$-band and of the $(x^2-y^2)$ one, or of a band made of any
linear combinations thereof of the type~(1), should coincide.)

But according to the double-exchange model electrons can move only
if localized moments ($t_{2g}$-spins) of the corresponding
sites are ordered ferromagnetically (although without doping, in
CaMnO$_3$ ($x=1$), the magnetic ordering is antiferromagnetic
(G-type)). Now, if we make the band e.g.\ out of
($x^2-y^2$)-orbitals, the band dispersion would have the form
\begin{equation}
\epsilon(\vec k)=-2t(\cos k_x+\cos k_y)
\end{equation}
i.e.\ the electrons in this band move only in the $xy$-plane, but
there is no dispersion in the $z$-direction. Therefore to gain full
kinetic energy it is enough to make this plane ferromagnetic, and the
adjacent planes may well remain antiparallel to the first one. But
this is just the A-type magnetic structure (ferromagnetic planes
stacked antiferromagnetically).
In this state we gain the same energy as in a fully ferromagnetic state
(because the position of the bottom of the band is the same),
but lose less exchange energy of localized $t_{2g}$-electrons, because
two out of six bonds for each Mn are still antiferromagnetic. The same
applies also to the C-type ordering, where four out of six bonds are
antiferromagnetic. This explains why in the case of double exchange via
degenerate orbitals --realized in overdoped manganites-- predominantly
these partly ferromagnetic (A-type and C-type) occur instead of 
full ferromagnetic ordering.

Note that in this case the electron occupy predominantly
$(x^2-y^2)$-states (or $3z^2-r^2$-states in case of C-type ordering).
Accordingly there will be a corresponding lattice distortion
(compression along $c$-axis, $c/a<1$, for the A-type structure, and
$c/a>1$ for the C-type one). But we want to stress that these are,
strictly speaking, not the localized orbitals, but rather the {\it bands}
of corresponding character.  Whether we should call it orbital
ordering, is a matter of convention (usually this terminology is
applied to the case of localized orbitals).  In any case, the feature
mentioned above (\cite{kanamori,khomskii2p}), that due to higher-order
effects, in particular lattice anharmonicity, only locally elongated
MeO$_6$-octahedra are observed in practice, is valid only for orbital
ordering of {\it localized} orbitals, and it is in general not true
for the band situation considered here.

Thus the double exchange via degenerate orbitals may quite naturally
lead to anisotropic magnetic structures (A-type or C-type): we gain
by that the full kinetic energy without being forced to sacrifice all
the exchange interaction of localized electrons (part of the bonds
remain antiferromagnetic). Which particular state will be stable at
which part of the phase diagram, is determined by the competition
between these terms, kinetic energy versus exchange energy, with the
electron energy depending on the band filling and sensitive to the density of
states for the corresponding band.

There are several factors which can complicate this picture. Thus, one
may in principle get in this case canted states, and not the fully
saturated A- or C-type structures~\cite{vandenbrink2,soloviev2}. There may also
appear inhomogeneous phase-separated states. The possibility of the
charge ordering (e.g.\ in the form of stripes) was also not considered
in~\cite{vandenbrink2}. But altogether it shows that the conventional
double-exchange picture should be modified if double-exchange goes
via degenerate orbitals, and the overall tendency which results due to
this is that not the simple ferromagnetic state, but more complicated
magnetic structures may be stabilized, which agrees with the general
tendency observed in experiment. This factor may be important in
explaining the strong qualitative asymmetry of the phase diagram of
manganites for  $x < 0.5$ (underdoped) and $x > 0.5$ (overdoped)
regimes.

\begin{figure}
\centering  \includegraphics[angle=90,totalheight=40mm]{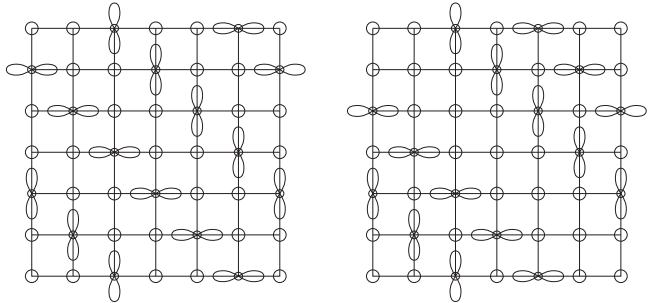}
\caption{
Left: single stripes (``Wigner crystal''). Right:
paired stripes, or bistripes in La$_{1-x}$Ca$_x$MnO$_3$ for $x=\frac23$.  
Mn$^{4+}$ ions denoted by O and Mn$^{3+}$ ions by $8$, $\infty$. 
}
\label{Fig:stripes}
\end{figure}

It is very interesting that stripes (and possibly bistripes) were
observed in overdoped manganites~\cite{radaelli,mori}, notably in 
La$_{1-x}$Ca$_x$MnO$_3$ for $x=\frac{2}{3}$ and  $x=\frac{3}{4}$.
Orbital degrees of freedom seem to play an important role in their
formation as well. The structure of the stripe and bistripe phases, 
see Fig.~\ref{Fig:stripes}, resembles somewhat that of the CE-phase at $x=1/2$.

The physical mechanism of the stripe formation in manganites could
in principle be similar to that involved in the cuprates~\cite{Zaanen},
i.e. coupling between charge and spin degrees of freedom. In manganites,
however, stripes are usually formed at temperatures above those of
magnetic ordering. As we see from Fig.~\ref{Fig:stripes}, stripes in this case
also imply a specific orbital ordering. We may think that orbital degrees
of freedom and the corresponding lattice distortions
strongly contribute to the very formation of stripes. One can indeed
show that the states with a particular orbital orientation attract another,
which can provide a mechanism for stripe formation~\cite{khomskii4}.
As a result, depending on the values of parameters, both the single
and paired stripe phases may be stabilized in overdoped manganites due to
this effect~\cite{khomskii3} 
that heavily relies on the presence of orbital degrees of freedom. 

\section{Conclusions}
\label{Sec:conclusions}
In conclusion we can only repeat that orbital effects play a very
important role in the physics of manganites, and also in many other
transition metal oxides. Together with
charge and spin degrees of freedom they determine all the  rich
variety of the properties of manganites in different doping regions.
Orbital effects also play a very important role in disordered phases,
determining to a large extent their transport and other properties.

An important recent achievement in this field is the development of
a method to directly study orbital ordering using 
resonant X-ray scattering~\cite{murakami}. This method was successfully
applied to a number of problems in
manganites~\cite{endoh,zimmermann,Murakami98,Wakabagashi00} 
as well as to several other systems. And although there is still a controversy
as to the detailed microscopic explanation of these
observations~\cite{ishihara2,elfimov,Benfatto99,benedetti}, this method will be
definitely of great use in the future.

Another interesting new development is the observation
of orbital excitations in LaMnO$_3$. And although many questions
here still remain unclear, the recent experimental progress 
will definitely open a new chapter in the study of orbital effects in oxides, in
particular in manganites. 

In our review we left out several effects for which orbital degrees of freedom
also may play an important role, e.g. phase separation~\cite{Dagotto01}
or short-range correlations above the ferromagnetic ordering 
temperature~\cite{Vasiliu99,Adams00,Dai00}. All these problems are
now under active investigation, and the results will definitively shed new light
on the physics of manganites, including the phenomenon of colossal magneto-resistance.
In summary, we see that the field of orbital physics is still
capable of producing important new results, and sometimes --surprises.

\section{Acknowledgments}
 
The authors gratefully acknowledge the collaboration with 
L.F. Feiner, P. Horsch, R. Kilian, K. Kugel, M. Mostovoy, A. M. Ole\'s, 
and G.A. Sawatzky and we also thank them for fruitful discussions.

%%%%%%%%%%%%%%%%% References %%%%%%%%%%%%%%%%%%

%\begin{references}


\begin{thebibliography}{10pt}


%%%%%%% Review articles:


\bibitem{goodenough} J.~B.~Goodenough, Magnetism and
Chemical Bond, Interscience Publ., New~York--London, 1963

\bibitem{kugel} K.~I.~Kugel and D.~I.~Khomskii, Sov.\ Phys.\ Usp.\
{\bf 25}, 231 (1982).

\bibitem{nagaosa} Y.~Tokura and N.~Nagaosa, Science {\bf 288}, 462 (2000).

\bibitem{oles00} A. M. Ole\'s, M. Cuoco and N. B. Perkins, in {\it Lectures on the Physics of 
Highly Correlated Electron Systems IV}, ed. F. Mancini, 
AIP Conf. Proc. No. 527 (AIP, New York, 2000).

\bibitem{jahn} H.~J.~Jahn and E.~Teller, Proc. Roy. Soc. A{\bf 161}, 220 (1937).

\bibitem{khomskii} D.~Khomskii and G.~Sawatzky, Solid State Comm. {\bf 102}, 87 (1997).

\bibitem{khomskii2} D.~Khomskii, 
"Electronic structure, exchange and magnetism in oxides", in {\it Spin Electronics} p. 89, 
eds. M. Ziese and M.J. Thornton, Springer Verlag (2001).

\bibitem{wollan} E.~O.~Wollan and W.~C.~Koehler, Phys. Rev. {\bf 100}, 545 (1955).

\bibitem{jirak} Z.~Jirak {\it et al.}, J. Magn. Magn. Mater. {\bf 53}, 153 (1985).

\bibitem{zimmermann} M.~v.~Zimmermann {\it et al.}, Phys. Rev. Lett.  {\bf 83}, 4872 (1999).

\bibitem{klingerer} T.~Niem\"oller et al, Eur. Phys. J. {\bf B8}, 5 (1999); 
R.~Klingerer {\it et al.}, preprint (2001).

\bibitem{endoh} Y.\ Endoh, K.\ Hirota, S.\ Ishihara, S.\ Okamoto, 
Y.\ Murakami, A.\ Nishizawa, T.\ Fukuda, H.\ Kimura, H.\ Nojiri, 
K.\ Kaneko, and S.\ Maekawa, Phys.\ Rev.\ Lett.\ {\bf 82}, 4328
(1999).

\bibitem{degennes} P.~G.~de~Gennes, Phys. Rev. {\bf 118}, 141 (1960).

\bibitem{radaelli} P.~G.~Radaelli {\it et al.}, Phys. Rev. B {\bf 59}, 14440 (1999).

\bibitem{mori} S.~Mori, C.~H.~Chen and S.-W.~Cheong, Nature {\bf 392}, 479
(1998).

\bibitem{wang} R.~Wang {\it et al.}, Phys. Rev. B {\bf 61}, 11946 (2000).

\bibitem{martin} C.~Martin {\it et al.}, J. Solid State Chem. {\bf 134}, 198 (1997).

\bibitem{vandenbrink2}J.~van~den~Brink and D.~Khomskii, Phys. Rev.
Lett. {\bf 82}, 1016 (1999).

\bibitem{gehring} G.~A.~Gehring and K.~A.~Gehring, Rep. Progr. Phys. {\bf 38}, 1
(1975).

\bibitem{kugel2} K.~I.~Kugel and D.~Khomskii, JETP Lett. {\bf 15}, 446 (1972); 
K.~I.~Kugel and D.~Khomskii, Sov.~Phys.--JETP {\bf 37}, 725 (1973).

\bibitem{anderson} P.~W.~Anderson, Phys. Rev. {\bf 115}, 2 (1959).

\bibitem{kanamori} J.~Kanamori, J. Appl. Phys. (Suppl.) {\bf 31},14S (1960).

\bibitem{khomskii2p} D.~Khomskii and J.~van~den~Brink, Phys. Rev. Lett.  {\bf 85}, 
3229 (2000).

\bibitem{halperin} R.~Englman and B.~Halperin, Phys. Rev. B {\bf 2}, 75 (1970); 
B.~Halperin and R.~Englman, Phys. Rev. B {\bf 3}, 1698 (1971).

\bibitem{eshelby} J.~D.~Eshelby, Solid State Phys., ed.~F.~Seitz and
D.~Turnbull, Academic Press, New York, v.~3, p.~79 (1956).

\bibitem{khachaturyan} A.~G.~Khachaturyan, Theory of Phase
Transformations and the Structure of Solid Solutions, Nauka, Moscow (1974).

\bibitem{khomskii3} D.~Khomskii and K.~I.~Kugel, Europhys. Lett. {\bf 55}, 208  (2001). 

\bibitem{khomskii4} D.~Khomskii and K.~I.~Kugel, cond-mat/0112340  (2001). 

\bibitem{yamada} Y.~Yamada {\it et al.}, Phys. Rev. Lett.  {\bf 77}, 904 (1996).

\bibitem{Inami99} T. Inami {\it et al.}, Jap. J. Appl. Phys., Suppl {\bf 38-1}, 212 (1999).

\bibitem{Yamada00} Y. Yamada {\it et al.}, Phys. Rev. B. {\bf 62}, 11600 (2000).

\bibitem{Cox01} D. Cox {\it et al.}, Phys. Rev. B. {\bf 64}, 024431 (2001).

\bibitem{zimmermann2} M.~v.~Zimmermann {\it et al.}, cond-mat/0007321 (2000).

\bibitem{mizokawa} T.~Mizokawa, D.~Khomskii and G.~Sawatzky, Phys. Rev. B {\bf 61}, R3776 (2000).

\bibitem{mizokawa2} T.~Mizokawa, D.~Khomskii and G.~Sawatzky, Phys. Rev. B {\bf 63}, 024403 (2000).

\bibitem{khaliullin} R.~Kilian and G.~Khaliullin, Phys. Rev. B {\bf 60}, 13458 (1999).

\bibitem{louca} D.~Louca and T.~Egami, J. Appl. Phys. {\bf 81}, 5484 (1997); 
Phys. Rev. B. {\bf 59}, 6193 (1999).

\bibitem{ishihara97} S.~Ishihara, M.~Yamanaka and N.~Nagaosa, Phys. Rev. B {\bf 56}, 
6861 (1997).

%\bibitem{oudovenko} G.~Khaliullin and V.~Oudovenko, Phys.Rev. B {\bf 56}, R14243 
%(1997)

%\bibitem{feiner} L.~F.~Feiner, A.~M.~Oles and J.~Zaanen, J. Phys. Condens. Matter
%{\bf 10}, L555 (1998)

\bibitem{khomskii5} D.~Khomskii, cond-mat/0004034 (2000).

\bibitem{vandenbrink3} J.~van~den~Brink and D.~Khomskii, Phys. Rev. B {\bf 63}, 1401416(R), (2001).

\bibitem{maezono} R.~Maezono and N.~Nagaosa, Phys. Rev. B {\bf 62}, 11576 (2000).

\bibitem{takahashi} A.~Takahashi and H.~Shiba, J. Phys. Soc. Jap. {\bf 69}, 3328 (2000).

\bibitem{ishihara3} S.~Ishihara, J.~Inoue and S.~Maekawa, Phys. Rev. B {\bf 55}, 8280 (1997).

\bibitem{vandenbrink4} J.~van~den~Brink {\it et al.}, Phys. Rev. B {\bf 58}, 10276 (1998).

\bibitem{KHA97} G.~Khaliullin and V.~Oudovenko, Phys. Rev. B 
{\bf 56}, R14243 (1997).

\bibitem{brink01} J.~van~den~Brink, Phys. Rev. Lett. {\bf 87}, 217202 (2001).

%%%%%%%%%%%% Orbiton experiment:

\bibitem{saitoh} E.~Saitoh, S.~Okamoto, K.~T.~Takahashi, K.~Tobe,
K.~Yamomoto, T.~Kimura, S.~Ishihara, S.~Maekawa, and Y.~Tokura, 
Nature (London) {\bf 410}, 180 (2001).
 
\bibitem{Allen01} P.B. Allen and V. Perebeinos, Nature {\bf 410}, 155 (2001).

\bibitem{FEI97}   L.~F.~Feiner, A.~M.~Ole\'s,  and J.~Zaanen,
Phys.\ Rev.\ Lett.\ {\bf 78}, 2799 (1997).

\bibitem{Perebeinos00} V. Perebeinos and P.B. Allen, Phys. Rev. Lett. {\bf 85}, 5178 (2000).
%%%%%%%%%%%%%%% Orbital polarons:

\bibitem{KIL99} R.~Kilian and G.~Khaliullin, Phys.\ Rev.\ B {\bf 60}, 
13458 (1999). 

%%%%%%%%%%%%%%% Orbital t-J model:
\bibitem{BHO00} J. van den Brink, P. Horsch and A.M. Ole\'s 
Phys. Rev. Lett. {\bf 85}, 5174 (2000). 

\bibitem{GOO55} J.~B.~Goodenough, Phys.\ Rev.\ {\bf 100}, 564 (1955).

%\bibitem{KAN59} J.~Kanamori, J.\ Phys.\ Chem.\ Solids {\bf 10}, 87 (1959).


%\bibitem{TOK00} Y.~Tokura and N.~Nagaosa, Science {\bf 288},462 (2000).

%%%%%%%%%%%% Gap in Kugel-Khomskii model:

%\bibitem{FEI99} For a physically realistic superexchange model in manganites
%see L.~F.~Feiner, A.~M.~Ole\'s, Phys.\ Rev.\ B {\bf 59}, 3295 (1999).

\bibitem{Villain80} J. Villain {\it et al.}, J. Physique {\bf 41}, 1263 (1980).

\bibitem{SHE82} E.F. Shender, Sov. Phys.--JETP {\bf 56}, 178 (1982).

\bibitem{TSV95} For a discussion of the order from disorder phenomena
in frustrated spin systems, see A.~M.~Tsvelik,
{\it Quantum Field Theory in Condensed Matter Physics}
(Cambridge University Press, Cambridge, 1995), Chap.~17,
and references therein.

\bibitem{FEI98}   L.~F.~Feiner, A.~M.~Ole\'s,  and J.~Zaanen,
J. Phys. Condens.\ Matter\ {\bf 10}, L555 (1998).

\bibitem{KHA99} G.~Khaliullin and R.~Kilian,
J. Phys. Condens.\ Matter\ {\bf 11}, 9757 (1999).

%%%%%%%%%%%%%%%% Frustration:-orbital model in ferro case:

\bibitem{BRI99} J.\ van den Brink, P.\ Horsch, F.\ Mack, and A.\ M.\ Ole\'s, 
Phys.\ Rev.\ B {\bf 59}, 6795 (1999). 

%%%%%%%%%%%%%%% Frustration:-cubic model:

\bibitem{NOTE1} See the discussion in Ref. \cite{kugel}, p.253.

\bibitem{KHA01} G.~Khaliullin,  Phys.\ Rev.\ B {\bf 64}, 212405 (2001).

%\bibitem{BEL76} E. Belorizky, R. Casalegno ans P. Fries, Phys. Stat. Sol. b 
%			{\bf 77}, 495 (1976).
 
%%%%%%%%%%%%% Orbital model in ferro:

\bibitem{KHA02} G.~Khaliullin and P.~Horsch, unpublished.

\bibitem{KUB02} K. Kubo, J. Phys. Soc. Jpn. 71 (2002), in press.

\bibitem{GLA73} S.J. Glass and J.O. Lawson, Phys. Lett. {\bf A 43} 234 (1973).

%%%%%%%%%%%%% Frustration: t_2g systems:

\bibitem{KUG75} K.I. Kugel and D.K. Khomskii, Sov. Phys.-Solid State Phys. 
{\bf 17} 285 (1975).

\bibitem{NOTE2} Note also the weaker orbital-lattice coupling and hence
the less destructive effect of cooperative lattice distortions
on orbital fluctuations in $t_{2g}$-systems.

\bibitem{KHA00a} G.~Khaliullin, and S.~Maekawa,
Phys.\ Rev.\ Lett.\ {\bf 85}, 3950 (2000).

\bibitem{KEI00} B.~Keimer, D.~Casa, A.~Ivanov, J.~W.~Lynn, M.v.~Zimmermann,
J.~P.~Hill, D.~Gibbs, Y.~Taguchi, and Y.~Tokura, 
Phys.\ Rev.\ Lett.\ {\bf 85}, 3946 (2000).

\bibitem{KHA01a} G.~Khaliullin, P.~Horsch, and A.~M.~Ole\'s,
Phys.\ Rev.\ Lett.\ {\bf 86}, 3879 (2001).

\bibitem{ISH97a} S.\ Ishihara, J.\ Inoue, and S.\ Maekawa, Phys.\ Rev.\ B 
{\bf 55}, 8280 (1997).

\bibitem{Bulaevskii68} L.N. Bulaevskii, E.L. Nagaev and D.I. Khomskii,
Sov. Phys.--JETP {\bf 27}, 836 (1968).

\bibitem{Uhlenbruck99} S. Uhlenbrusk {\it et al.}, Phys. Rev. Lett. {\bf 82}, 185 (1999).

%%%%%%%%%%%%% Ferro-insulator(exp):

%%%%%%%%%%%%% Polaron spin moment(exp):

\bibitem{TER97} J.\ M.\ De Teresa, M.\ R.\ Ibarra, P.\ A.\ Algarabel, 
C.\ Ritter, C.\ Marquina, J.\ Blasco, J.\ Garcia, A.\ del Moral, and 
Z.\ Arnold, Nature (London) {\bf 386}, 256 (1997).

%%%%%%%%%%%%%%% DE + lattice polarons:

\bibitem{MIL96} A.\ J.\ Millis, B.\ I.\ Shraiman, and
R.\ Mueller, Phys.\ Rev.\ Lett.\ {\bf 77}, 175 (1996);
A.\ J.\ Millis, Nature (London) {\bf 392}, 147 (1998).

\bibitem{ROD96} J.\ H.\ R\"oder, J.\ Zang, and A.\ R.\ Bishop,
Phys.\ Rev.\ Lett.\ {\bf 76}, 1356 (1996).

%%%%%%%%%%%% Phase separation:

\bibitem{MOR99} For a discussion of phase separation issue, see, 
for example, A.~Moreo, S.~Yunoki, and E.~Dagotto, 
Science {\bf 283}, 2034 (1999).

%%%%%%%%%%%%%%% DE model:

\bibitem{ZEN51} C.\ Zener, Phys.\ Rev.\ {\bf 82}, 403 (1951).

\bibitem{AND55} P.\ W.\ Anderson and H.\ Hasegawa, 
Phys.\ Rev.\ {\bf 100}, 675 (1955).

%\bibitem{DEG60} P.-G.\ de Gennes, Phys.\ Rev.\ {\bf 118}, 141 (1960).

\bibitem{KUB72} K.\ Kubo and N.\ Ohata, J.\ Phys.\ Soc.\ Jpn.\ 
{\bf 33}, 21 (1972).

%%%%%%%%%%%%%%%% Optical spectra in ferro:

\bibitem{OKI95} Y.\ Okimoto, T.\ Katsufuji, T.\ Ishikawa, A.\ Urushibara, 
T.\ Arima, and Y.\ Tokura, Phys.\ Rev.\ Lett.\ {\bf 75}, 109 (1995).

\bibitem{SHI97} H.\ Shiba, R.\ Shiina, and A.\ Takahashi, 
J.\ Phys.\ Soc.\ Jpn. {\bf 66}, 941 (1997).

\bibitem{KIL98} R.\ Kilian and G.\ Khaliullin, Phys.\ Rev.\ B 
{\bf 58}, R11\,841 (1998).

\bibitem{FUR96} N.\ Furukawa, J.\ Phys.\ Soc.\ Jpn.\
{\bf 65}, 1174 (1996).

%%%%%%%%%%%%%%%% Magnon softening:

\bibitem{PER96} T.\ G.\ Perring, G.\ Aeppli, S.\ M.\ Hayden, 
S.\ A.\ Carter, J.\ P.\ Remeika, and S.-W.\ Cheong, Phys.\ Rev.\ Lett.\ 
{\bf 77}, 711 (1996). 

\bibitem{HWA98} H.\ Y.\ Hwang, P.\ Dai, S.-W.\ Cheong, 
G.\ Aeppli, D.\ A.\ Tennant, and H.\ A.\ Mook, Phys.\ Rev.\ Lett. {\bf 80}, 
1316 (1998).

\bibitem{KHA00} G.~Khaliullin and R.~Kilian, Phys.\ Rev.\ B {\bf 61}, 
3494 (2000).

\bibitem{stekelen} J. van den Brink, W. Stekelenburg, D.I. Khomskii, 
G.A. Sawatzky and K.I. Kugel, Phys. Rev. B. {\bf 58}, 10276 (1998)

\bibitem{FER98} J.\ A.\ Fernandez-Baca, P.\ Dai, H.\ Y.\ Hwang, 
C.\ Kloc, and S.-W.\ Cheong, Phys.\ Rev.\ Lett.\ {\bf 80}, 4012 (1998).

%\bibitem{Wollan55} E.O. Wollan and W.C. Koehler, Phys. Rev. {\bf 100}, 545 (1955).

\bibitem{Kuwahara95} H. Kuwahara {\it et al.}, Science {\bf 270}, 961  (1995).

\bibitem{Kawano97} H. Kawano {\it et al.}, Phys. Rev. Lett.  {\bf 78}, 4253 (1997).


\bibitem{Tomioka96} Z. Jirac {\it et al.}, J. Magn. Magn. Mater. {\bf 53}, 153 (1985),
                        Y. Tomioka {\it et al.}, Phys. Rev. B  {\bf 53},  R1689 (1996).

\bibitem{Urushibara95}  Y. Okimoto {\it et al.}, Phys. Rev. Lett.  {\bf 75}, 109 (1995),
                        P. Schiffer {\it et al., ibid.}   {\bf 75}, 3336 (1995),
                        P. G. Radaelli {\it et al., ibid.}  {\bf 75}, 4488 (1995),
                        Y. Okimoto {\it et al.}, Phys. Rev. B  {\bf 55}, 4206 (1997).

\bibitem{Mori98} S. Mori, C.H. Chen and  S.-W. Cheong, Nature {\bf 392}, 473 (1998).

\bibitem{Kimura_prep} T. Kimura {\it et al.}, unpublished.

\bibitem{Moritomo95}    Y. Moritomo {\it et al.},  Phys. Rev. B {\bf 51},  3297 (1995),
                        B.J. Sternlieb {\it et al.}, Phys. Rev. Lett. {\bf 76}, 2169 (1996),
                        Y. Moritomo  {\it et al.}, Nature  {\bf380}, 141 (1996).


\bibitem{Tokunaga98} M. Tokunaga  {\it et al.},  Phys. Rev. B {\bf 57},  5259 (1998).

\bibitem{soloviev} I.~V.~Solovyev and K.~Terakura, Phys. Rev. Lett. {\bf 83}, 2825 (1999).

\bibitem{vandenbrink} J.~van~den~Brink, G.~Khaliullin and D.~Khomskii, Phys. Rev. Lett. 
{\bf 83}, 5118(1999).

\bibitem{Babushkina99} N.A. Babushkina {\it et al.}, Phys. Rev. B {\bf59}, 6994 (1999), 
        M. Uehara {\it et al.}, Nature {\bf 399}, 560 (1999).

\bibitem{Kagan01} M.Y. Kagan, D.I. Khomskii and K.I. Kugel, JETP {\bf 93}, 415 (2001).

\bibitem{commentCE} S.-Q. Shen, Phys. Rev. Lett. {\bf 86}, 5842 (2001).

\bibitem{replyCE} J. van den Brink, G. Khaliullin and D. Khomskii, Phys. Rev. Lett. 
{\bf 86}, 5843 (2001).

\bibitem{kuwahara} H.~Kuwahara {\it et al.}, Mat. Res. Soc. Symp. Proc.,
v.~494, 83 (1998).

\bibitem{soloviev2} I.~Solovyev and K.~Terakura, Phys. Rev. B in press (2001).

\bibitem{Zaanen} J. Zaanen and O. Gunnarsson, Phys. Rev. B {\bf 40}, 7391 (1989).

\bibitem{murakami} Y. Murakami {\it et al.}, Phys. Rev. Lett.  {\bf 80}, 1932 (1998).

\bibitem{Murakami98} Y. Murakami {\it et al.}, Phys. Rev. Lett.  {\bf 81}, 582 (1998).

\bibitem{Wakabagashi00} Y. Wagabagashi {\it et al.}, J. Phys. Soc. Jap. {\bf 69}, 2731 (2000).

\bibitem{ishihara2} S.~Ishihara and S.~Maekawa, Phys. Rev. Lett.  {\bf 80}, 3799 (1998).

\bibitem{elfimov} I.~S.~Elfimov, V.~A.~Anisimov and G.~Sawatzky, Phys. Rev. Lett.  {\bf 82}, 4264
(1999).

\bibitem{Benfatto99} M. Benfatto {\it et al.}, Phys. Rev. Lett. {\bf 83}, 636 (1999).

\bibitem{benedetti} P. Benedetti, J. van den Brink, E. Pavarini, A. Vigliante, and P. Wochner, 
Phys. Rev. B. {\bf 63}, 60408(R) (2001).

\bibitem{Dagotto01} E. Dagotto, T. Hotta and A. Moreo, Phys. Rep. {\bf 344}, 1 (2001).

\bibitem{Vasiliu99}  L. Vasiliu-Doloc {\it et al.}, Phys. Rev. Lett. {\bf 83}, 4393 (1999).

\bibitem{Adams00}  C.P. Adams {\it et al.}, Phys. Rev. Lett. {\bf 85}, 3954 (2000).

\bibitem{Dai00}  P. Dai {\it et al.}, Phys. Rev. Lett. {\bf 85}, 2553 (2000).

\end{thebibliography}
\end{document}